\documentstyle[12pt,axodraw]{article}

\begin{document}
\baselineskip=17pt
\parskip=5pt

\begin{titlepage} \footskip=5in     

\title{ 
\begin{flushright} \normalsize 
Fermilab-Pub-00/137-T \vspace{0.1em} \\ 
June 2000 \vspace{5em} \\  
\end{flushright}
\large\bf   
Reanalysis of  $\boldmath CP$  Violation in  
$\,K_{\rm L}^{}\rightarrow\pi^+\pi^-\gamma\,$}  
   
\author{\normalsize\bf 
Jusak~Tandean$^a$\thanks{E-mail address: jtandean@iastate.edu} \ and  
{G.~Valencia$^{a,b}$\thanks{E-mail address: valencia@iastate.edu}} \\
\normalsize\it 
$^a$Department of Physics and Astronomy, Iowa State University, 
Ames, IA 50011 \\  
\normalsize\it 
$^b$Fermi National Accelerator Laboratory, P.O.Box 500, Batavia, IL 60510
}    
    
\date{}   
\maketitle   
   
\begin{abstract} 
  
We present a reanalysis of direct $CP$ violation in the decay 
$\,K_{\rm L}^{}\rightarrow\pi^+\pi^-\gamma\,$. 
We point out an existing discrepancy between the theoretical 
and experimental definitions of $\epsilon'_{+-\gamma}$. 
Adopting the experimental definition of  $\epsilon'_{+-\gamma}$, 
we estimate that $\epsilon'_{+-\gamma}/\epsilon$ could be as 
large as a few times~$10^{-4}$ both within the standard model 
and beyond. We discuss these estimates in detail and we also 
show how a judicious choice of $E_\gamma^*$ cuts can increase 
the sensitivity of the observable $\epsilon'_{+-\gamma}$ to 
the underlying $CP$ violation.

\end{abstract}   
   
\end{titlepage}

\section{Introduction}

The origin of  $CP$  violation remains an unsolved problem 
in particle physics.  
An important piece of this puzzle has been provided recently by 
the KTeV and NA48 collaborations~\cite{e'x} that experimentally 
confirmed the presence of direct $CP$ violation 
in the  $\,K\rightarrow\pi\pi\,$ neutral-kaon decays.     
Interestingly, the measured results were larger than 
most existing standard model estimates at the time.  The difference 
could be attributed to new physics, although updated analyses 
of theoretical uncertainties indicate that there is no 
serious disagreement with the standard model~\cite{e't}.    
This situation underscores the importance of observing $CP$ 
violation in other reactions.

The decays  $\,K_{\rm L,S}^{}\rightarrow\pi^+\pi^-\gamma\,$   
have long been 
recognized~\cite{SehWol,CosKab,DonHV,LitVal,DamIsi2} 
for their potential to test $CP$ violation. 
The amplitude for each of these decays is conventionally  
divided into two contributions:       
inner bremsstrahlung~(IB) and direct emission~(DE).  
The IB is completely determined by the 
$\,K\rightarrow\pi^+\pi^-\,$  process that underlies it,  
whereas the DE part encodes additional dynamical features. 
The  $\,K_{\rm S}^{}\rightarrow\pi^+\pi^-\gamma\,$  decay 
is known to be dominated by the IB, but in the  $K_{\rm L}^{}$  
decay the IB is suppressed because the underlying  
$\,K_{\rm L}^{}\rightarrow\pi^+\pi^-\,$  process is  
$CP$-violating.  
This makes the more interesting DE term in the $K_{\rm L}^{}$  
decay  more accessible, thereby raising the possibility of 
observing  new direct  $CP$  violation in this mode.

The presence of  $CP$  violation in  
$\,K_{\rm L}^{}\rightarrow\pi^+\pi^-\gamma\,$  
has been observed in recent experiments~\cite{eta+-g1,eta+-g2}, 
and this is characterized by the parameter  $\eta_{+-\gamma}$, 
defined in analogy to the parameter  $\eta_{+-}^{}$  
in the  $\,K\rightarrow\pi^+\pi^-\,$  case.       
However, the result has not yet reached the level of sensitivity 
needed for detecting direct  $CP$  violation in 
$\,K_{\rm L}^{}\rightarrow\pi^+\pi^-\gamma.\,$    
On the theoretical side, studies on direct $CP$  violation in 
this decay have been performed by various 
authors~\cite{DonHV,DamIsi2,LinVal,various,ColIP}.   
In this paper, we revisit this subject for the following 
reasons. First, the existent experimental analysis of the  $CP$-odd 
observable  $\eta_{+-\gamma}$ is not consistent with the 
theoretical definitions because 
it assumes that this quantity is a constant over phase space. 
Second, KTeV will perform a new analysis of this mode with 
their recent data and it is therefore timely to 
update the theoretical expectations both within and beyond 
the standard model. Finally, we  examine the possibility 
of measuring new, direct, $CP$ violation in this mode 
from an analysis of the decay distribution~\cite{deff3}  
without $K_{\rm S}^{}-K_{\rm L}^{}$ interference.

In the following section we introduce our notation for 
the relevant decay amplitudes and the  $CP$-violating 
parameters. In particular, we introduce the parameter, 
$\hat\epsilon$, to characterize the {\it new} direct 
$CP$ violation in these modes. 
In Section~\ref{int}, we relate these parameters 
to $CP$-violating observables occurring in the interference  
between the amplitudes for  $K_{\rm L}^{}$  and  
$K_{\rm S}^{}$  decays into  $\pi^+\pi^-\gamma$.    
By carefully treating the energy dependence in 
the amplitudes, we derive a relation between 
$\hat\epsilon$, and the experimental observable 
$\epsilon'_{+-\gamma}$. In this manner we extract the 
current limits on $\hat\epsilon$. 
In Section~\ref{IBDEE}, we consider detecting  direct  
$CP$  violation in the interference between the inner 
bremsstrahlung and direct emission in the electric 
amplitude of  
$\,K_{\rm L}^{}\rightarrow\pi^+\pi^-\gamma.\,$   
Using a recent KTeV result, we extract a bound  
on  $\hat\epsilon$. 
In Section~\ref{SM} we present an estimate for  $\hat\epsilon$  
in the standard model. Finally,    
in Section~\ref{NP},  we consider potentially large 
new physics contributions to  $\hat\epsilon$ that 
may arise in left-right symmetric  models~\cite{mohapatra}  
and in generic supersymmetric models~\cite{GabGMS}.

\section{Amplitudes and Direct $\boldmath CP$-Violating 
Parameters\label{a}}

The amplitudes for  $\,K^0\rightarrow\pi^+\pi^-,\pi^0\pi^0\,$  
are conventionally written as  
\begin{eqnarray}   \label{K->2pi}   
{\cal M}_{K^0\rightarrow\pi^+\pi^-}^{}  \;=\;  
A_0^{}\, {\rm e}^{{\rm i}\delta_0^0}  
+ \mbox{$1\over\sqrt{2}$} A_2^{}\, {\rm e}^{{\rm i}\delta_0^2}   \;,  
\hspace{2em}  
{\cal M}_{K^0\rightarrow\pi^0\pi^0}^{}  \;=\;  
A_0^{}\, {\rm e}^{{\rm i}\delta_0^0}    
- \sqrt{2}\, A_2^{}\, {\rm e}^{{\rm i}\delta_0^2}   \;,   
\end{eqnarray}   
where $A_I^{}$ is the component for the  $\pi\pi$  state    
with isospin  $I$,  and   
$\delta_J^I$  is the strong-rescattering phase for 
the  $\pi\pi$  state with angular momentum  $J$.  
From  $\,K\rightarrow\pi\pi\,$  data, one can extract 
in the isospin limit 
$\,{\rm Re}\, A_0^{}\simeq2.72\times10^{-7}\,\rm GeV\,$     
and  
$\,\omega\equiv{\rm Re}\, A_2^{}/{\rm Re}\,
A_0^{}\simeq1/22.2\,$~\cite{pdb}.\footnote{Isospin violation 
significantly complicates this analysis \cite{GarVal2}.} 
The physical states  $K_{\rm L}^{}$  and  $K_{\rm S}^{}$  
are given by     
\begin{eqnarray}   
\bigl| K_{\rm L}^{} \bigr\rangle  \;=\;   
{ (1+\bar{\epsilon}) \bigl| K^0 \bigr\rangle 
 + (1-\bar{\epsilon}) \bigl| \bar{K}^0 \bigr\rangle 
 \over  \sqrt{2+2|\bar{\epsilon}|^2} }   \;,   
\hspace{3em}   
\bigl| K_{\rm S}^{} \bigr\rangle  \;=\;   
{ (1+\bar{\epsilon}) \bigl| K^0 \bigr\rangle 
 - (1-\bar{\epsilon}) \bigl| \bar{K}^0 \bigr\rangle 
 \over  \sqrt{2+2|\bar{\epsilon}|^2} }   \;,   
\end{eqnarray}   
where  $\bar{\epsilon}$ corresponds to  $CP$  
violation in the kaon-mass matrix~\cite{DonGH}  
and we use the convention 
$\,CP\bigl| K^0 \bigr\rangle= - \bigl| \bar{K}^0 \bigr\rangle .\,$   
The  $CP$-violating parameters in  
$\,K\rightarrow\pi\pi\,$ decays are  
\begin{eqnarray}   \label{e,e'}  
\epsilon  \;=\;  
\bar{\epsilon} + {{\rm i\,Im}\, A_0^{}\over{\rm Re}\, A_0^{}}   \;,  
\hspace{3em}   
\epsilon'  \;=\;  
{\omega\over\sqrt{2}} 
\left( {{\rm Im}\, A_2^{}\over{\rm Re}\, A_2^{}}   
      - {{\rm Im}\, A_0^{}\over{\rm Re}\, A_0^{}} \right) \,  
{\rm e}^{{\rm i}(\delta_0^2-\delta_0^0+\pi/2)}   \;,   
\end{eqnarray}   
corresponding to indirect and direct $CP$ violation, respectively. 
It is also conventional to define the ratio of amplitudes
\begin{eqnarray}   
\eta_{+-}  \;\equiv \;  
{ {\cal M}_{K_{\rm L}^{}\rightarrow\pi^+\pi^-}^{}  \over  
 {\cal M}_{K_{\rm S}^{}\rightarrow\pi^+\pi^-}^{} }    
\;=\;  \epsilon + \epsilon'   \;.      
\end{eqnarray}   
Experimentally it is found that 
$\;|\eta_{+-}^{}|=(2.285\pm0.019)\times 10^{-3}\,$ and that its 
phase  is $\,\phi_{+-}^{}=43.5^{\rm o}\pm0.6^{\rm o}\,$~\cite{pdb}.    
The recent measurements of direct $CP$ violation~\cite{e'x}   
lead to a new world average~\cite{na48}       
$\,\epsilon'/\epsilon=(19.3\pm2.4)\times 10^{-4}.\,$

In the  $\,K\rightarrow\pi\pi\gamma\,$  decay,  
the amplitude is generally decomposed into electric and 
magnetic terms.  
The electric part  $E$  receives contributions from both 
inner-bremsstrahlung and direct-emission processes, whereas 
the magnetic part  $M$  arises exclusively from the direct-emission.  
Our notation for the amplitude for 
$\,K\rightarrow\pi^+ \bigl( p_+^{} \bigr) \, 
\pi^- (p_-^{})\, \gamma(q),\,$  
where  $K$  is any neutral kaon  
$ \bigl( K^0, \bar{K}^0, K_{\rm L}, K_{\rm S}  \bigr) $, is
\begin{eqnarray}    
{\cal M}_{K\rightarrow\pi^+\pi^-\gamma}^{}  &=&  
\left[ E_{\rm IB}^{}(K) + E_{\rm DE}^{}(K) \right]     
{ \varepsilon_\mu^* 
 \left[ \bigl( p_+^{}+p_-^{} \bigr) {}^\mu \nu   
       - \bigl( p_+^{}-p_-^{} \bigr) {}^\mu z \right]  
 \over  m_K^{} }    
\nonumber \\ &&  
+\; 
M(K)\;   
{ 4\, \epsilon_{\lambda\mu\rho\sigma}^{}\,   
 \varepsilon^{\lambda*} p_+^\mu p_-^\rho q^\sigma  
 \over  m_K^3 }   \;,  
\end{eqnarray}   
with the kinematic variables    
\begin{eqnarray}   \label{nu,z}   
\nu  \;=\;  {2k\cdot \bigl( p_+^{}-p_-^{} \bigr) \over m_K^2}   \;,  
\hspace{2em}  
z  \;=\;  {2k\cdot q\over m_K^2}   \;.  
\end{eqnarray}   
In the kaon rest-frame, 
$\,\nu=2 \bigl( E_{\pi^+}^{}-E_{\pi^-}^{} \bigr) /m_K^{} \,$  
and  $\,z=2E_\gamma^*/m_K^{},\,$  with $E_\gamma^*$ being 
the usual notation for the photon energy in this frame.
  
For  $\,K^0\rightarrow\pi^+\pi^-\gamma,\,$   the IB term,   
\begin{eqnarray}   \label{IB}     
E_{\rm IB}^{} \bigl( K^0 \bigr)   \;=\;    
{ 4 e\, {\cal M}_{K^0\rightarrow\pi^+\pi^-}^{}  \over  
 m_K^{}\, (z^2-\nu^2) }   \;,  
\end{eqnarray}   
is completely  determined~\cite{EckNP}  by the amplitude 
for the underlying nonradiative decay  
$\,K^0\rightarrow\pi^+\pi^-.\,$  
Experimentally, it is separated by fitting the characteristic 
bremsstrahlung spectrum, which behaves as    
$\,{\rm d}\Gamma_{\rm IB}^{}/{\rm d}E_\gamma^*\sim1/E_\gamma^*\,$  
as  $\,E_\gamma^*\rightarrow0.\,$  
Following  Ref.~\cite{LitVal}, we write, guided by 
dimensional analysis, the DE terms       
\begin{eqnarray}    
E_{\rm DE}^{} \bigl( K^0 \bigr)  \;=\;   
e\, |G_8^{}|\, f_{\!\pi}^2\, \xi_{\rm E}^{}(\nu,z)   \;,  
\hspace{2em}   
M \bigl( K^0 \bigr)  \;=\;   
e\, |G_8^{}|\, f_{\!\pi}^2\, {\rm i}\xi_{\rm M}^{}(\nu,z)   \;,  
\end{eqnarray}   
so that the dimensionless form-factor 
$\xi_{\rm E}^{}$  $ \bigl( \xi_{\rm M}^{} \bigr) $  
in the electric (magnetic) amplitude 
is expected to be of order one. Also, 
$\,f_{\!\pi}^{}\simeq92.4\,\rm MeV\,$  
is the pion-decay constant  and  $G_8^{}$ is defined by    
\begin{equation}   \label{G8}   
A_0^{}  \;=\;   
\sqrt{2}\, G_8^{} f_{\!\pi}^{}\, \bigl( m_K^2-m_\pi^2 \bigr)   \;. 
\end{equation}  
The corresponding amplitudes for  
$\,\bar{K}^0\rightarrow\pi^+\pi^-\gamma\,$  are obtained 
by requiring $CPT$  invariance, and they are
\begin{eqnarray}   \label{IB'}  
\begin{array}{c}   \displaystyle  
E_{\rm IB}^{} \bigl( \bar{K}^0 \bigr)  \;=\;  
{ -4e\, {\cal M}_{K^0\rightarrow\pi^+\pi^-}^{*}  \over  
 m_K^{}\, (z^2-\nu^2) }   \;,  
\vspace{2ex} \\   \displaystyle  
E_{\rm DE}^{} \bigl( \bar{K}^0 \bigr)   \;=\;   
-e\, |G_8^{}|\, f_{\!\pi}^2\, \xi_{\rm E}^{*}(-\nu,z)    \;,  
\hspace{2em}   
M \bigl( \bar{K}^0 \bigr)   \;=\;   
+e\, |G_8^{}|\, f_{\!\pi}^2\, {\rm i} \xi_{\rm M}^{*}(-\nu,z)   \;,  
\end{array}   
\end{eqnarray}   
where in  ${\cal M}_{K\rightarrow\pi\pi}^{*}$  
and  $\xi_{\rm E,M}^{*}$  the complex conjugation refers  
only to the weak  $CP$-violating  phases 
and not to the strong final-state interaction phases.

The form factors  $\xi_{\rm E,M}^{}$  can be expressed in 
a multipole expansion~\cite{CosKab}.  There is almost no 
experimental information on the electric form-factor, as it 
generates small corrections to the amplitudes in both 
$K_{\rm S}$ and $K_{\rm L}$ decays. For this reason, it 
will be sufficient for us to assume that  $\xi_{\rm E}^{}$  is 
saturated by the leading multipole, E1,  and  use
\begin{eqnarray}   \label{xEDE}   
\xi_{\rm E}^{}(\nu,z)  \;=\;  
F_{\rm E}^{}\; {\rm e}^{{\rm i}\delta_1^1}   \;,     
\end{eqnarray}   
where  $F_{\rm E}^{}$  is a dimensionless complex constant, 
expected to be of order 1 by dimensional analysis. The final   
state interaction phase is $\delta_1^1$,  reflecting the fact 
that in this amplitude the two pions are in an  $\,I=J=1\,$  state. 
The magnetic form-factor, on the other hand, has been 
experimentally studied in some detail~\cite{deff1,deff2} and 
found to depend mostly on~$z$.   
For the remainder of this paper, we will take 
\begin{eqnarray}   \label{xMDE}   
\xi_{\rm M}^{}(\nu,z)  \;=\;  
\xi_{\rm M}^{}(z)\; {\rm e}^{{\rm i}\delta_1^1}   \;,   
\end{eqnarray}   
corresponding to the leading multipole, M1, with an $E_\gamma^*$ 
dependent form-factor. In what follows    
we neglect any  $\,\Delta I=3/2\,$  contribution 
to  $\xi_{\rm E,M}^{}$.

For the physical kaon states $K_{\rm S}^{}$  and  $K_{\rm L}^{}$  
the IB amplitudes are given by  
\begin{eqnarray}   \label{IB,K_S,K_L}   
E_{\rm IB}^{} \bigl( K_{\rm S}^{} \bigr)  \;=\;   
{ 4 e\, {\cal M}_{K_{\rm S}^{}\rightarrow\pi^+\pi^-}^{} 
 \over  m_K^{}\, (z^2-\nu^2)}   \;, 
\hspace{2em}  
E_{\rm IB}^{} \bigl( K_{\rm L}^{} \bigr)    
\;=\;  
\eta_{+-}^{}\, E_{\rm IB}^{} \bigl( K_{\rm S}^{} \bigr)   \;.   
\end{eqnarray}   
The electric DE amplitudes for 
$\,K_{\rm L,S}^{}\rightarrow\pi^+\pi^-\gamma\,$ are      
\begin{eqnarray}   \label{DEE}  
E_{\rm DE}^{} \bigl( K_{\rm L,S}^{} \bigr)  \;=\;  
e\, |G_8^{}|\, f_{\!\pi}^2\, \xi_{\rm E}^{\rm L,S}   \;,    
\end{eqnarray}   
where  
\begin{equation}   \label{xi_EL} 
\xi_{\rm E}^{\rm L}  \;=\;     
\sqrt{2}\, {\rm Re}\, F_{\rm E}^{}\, \left[ 
\epsilon  +  
{\rm i} 
\left( {{\rm Im}\, F_{\rm E}^{}\over {\rm Re}\, F_{\rm E}^{}}   
      - {{\rm Im}\, A_0^{}\over {\rm Re}\, A_0^{}} \right)  
\right] \, {\rm e}^{{\rm i}\delta_1^1}   \;,     
\end{equation}   
\begin{eqnarray}   \label{xi_ES} 
\xi_{\rm E}^{\rm S}  \;=\;  
\sqrt{2}\, {\rm Re}\, F_{\rm E}^{}\; {\rm e}^{{\rm i}\delta_1^1}   \;.       
\end{eqnarray}   
We have, as usual, dropped terms quadratic in weak phases.
Similarly, for the magnetic amplitudes we have 
\begin{eqnarray}   \label{DEM}  
M \bigl( K_{\rm L,S}^{} \bigr)  \;=\;  
e\, |G_8^{}|\, f_{\!\pi}^2\, {\rm i} \xi_{\rm M}^{\rm L,S}(z)   \;,    
\end{eqnarray}   
where  
\begin{equation}   \label{xi_ML} 
\xi_{\rm M}^{\rm L}  \;=\;     
\sqrt{2}\, {\rm Re}\, \xi_{\rm M}^{}\; 
{\rm e}^{{\rm i}\delta_1^1} 
\end{equation}   
\begin{eqnarray}   \label{xi_MS} 
\xi_{\rm M}^{\rm S}  \;=\;   
\sqrt{2}\, {\rm Re}\, \xi_{\rm M}^{}\, \left[  
\epsilon  +  
{\rm i} 
\left( { {\rm Im}\, \xi_{\rm M}^{}  \over  
        {\rm Re}\, \xi_{\rm M}^{} }   
      - {{\rm Im}\, A_0^{}\over {\rm Re}\, A_0^{}} \right)  
\right] \, {\rm e}^{{\rm i}\delta_1^1}   \;.       
\end{eqnarray}   

It is then possible to define two $CP$-violating quantities 
associated with the electric and magnetic amplitudes. 
For the electric amplitude, one has the usual  
ratio~\cite{DonHV,DamIsi2}\footnote{Notice that we call it 
$\tilde{\eta}_{+-\gamma}^{}$  instead of  $\eta_{+-\gamma}^{}$  
to differentiate it from the experimental observable.  
This is an important distinction that has not 
been appreciated in the literature.}
\begin{eqnarray}   \label{eta+-g}     
\tilde{\eta}_{+-\gamma}^{}  \;\equiv\;   
{ E_{\rm IB}^{} \bigl( K_{\rm L}^{} \bigr)    
 + E_{\rm DE}^{} \bigl( K_{\rm L}^{} \bigr)    
 \over   
 E_{\rm IB}^{} \bigl( K_{\rm S}^{} \bigr)    
 + E_{\rm DE}^{} \bigl( K_{\rm S}^{} \bigr) }     
\;=\;    
{ \eta_{+-}^{} + \displaystyle 
 \frac{|G_8^{}|\, f_{\!\pi}^2\, m_K^{}}{4\, {\rm Re}\, A_0^{}} 
 \frac{\xi_{\rm E}^{\rm L}}{\sqrt{2}} (z^2-\nu^2)\, 
 {\rm e}^{-{\rm i}\delta_0^0}    
 \over 
 1 + \displaystyle 
 \frac{|G_8^{}|\, f_{\!\pi}^2\, m_K^{}}{4\, {\rm Re}\, A_0^{}} 
 \frac{\xi_{\rm E}^{\rm S}}{\sqrt{2}} (z^2-\nu^2)\, 
 {\rm e}^{-{\rm i}\delta_0^0} }   \;.  
\end{eqnarray}   
Noting that  
\begin{eqnarray}   
\frac{|G_8^{}|\, f_{\!\pi}^2\, m_K^{}}{4\, A_0^{}} 
\;\simeq\;  \frac{f_{\!\pi}^{}}{4\sqrt{2}\, m_K^{}} 
\;\sim\;  0.03    
\end{eqnarray}   
and that  $\,z^2-\nu^2\,$  is always less than 1,  
we write  
\begin{eqnarray}   \label{teta+-g}     
\tilde{\eta}_{+-\gamma}^{}  &\simeq&   
\eta_{+-}^{} +   
{|G_8^{}|\, f_{\!\pi}^2\, m_K^{}\over4\, {\rm Re}\, A_0^{}} 
{ \bigl( \xi_{\rm E}^{\rm L}-\eta_{+-} \xi_{\rm E}^{\rm S} \bigr) 
 \over  \sqrt{2} } 
(z^2-\nu^2)\, {\rm e}^{-{\rm i}\delta_0^0}      
\nonumber \\ 
&\simeq&  
\eta_{+-} + 
{ |G_8^{}|\, f_{\!\pi}^2\, m_K^{}\, {\rm Re}\, F_{\rm E}^{}  
 \over  4\, {\rm Re}\, A_0^{} }\; 
{\rm i} \left( {{\rm Im}\, F_{\rm E}^{}\over{\rm Re}\, F_{\rm E}^{}} 
              - {{\rm Im}\, A_0^{}\over{\rm Re}\ A_0^{}} \right) \, 
(z^2-\nu^2)\, {\rm e}^{{\rm i}(\delta_1^1-\delta_0^0)}   \;.   
\end{eqnarray}   
It is, therefore, convenient to define the parameter    
\begin{eqnarray}   \label{he}     
\hat\epsilon  \;\equiv\;   
{ |G_8^{}|\, f_{\!\pi}^2 m_K^{}\, {\rm Re}\, F_{\rm E}^{}  
 \over  4\, {\rm Re}\, A_0^{} }     
\left( {{\rm Im}\, F_{\rm E}^{}\over{\rm Re}\, F_{\rm E}^{}}   
      - {{\rm Im}\, A_0^{}\over{\rm Re}\, A_0^{}} \right)   
\end{eqnarray}   
to characterize {\it new} direct  $CP$  violation in 
$\,K\rightarrow\pi^+\pi^-\gamma.\,$  
Thus, we have  
\begin{eqnarray}   \label{e_ppg}     
\tilde{\epsilon}_{+-\gamma}'  \;\equiv\;  
\tilde{\eta}_{+-\gamma}^{}  -  \eta_{+-}     
\;=\;  
\hat\epsilon\, {\rm e}^{{\rm i}\delta_\epsilon^{}}\, 
(z^2-\nu^2)   \;,  
\hspace{2em}  
\delta_\epsilon^{}  \;\equiv\;  \delta_1^1-\delta_0^0+\pi/2   \;.  
\end{eqnarray}   
The quantity  $\tilde{\epsilon}_{+-\gamma}'$  varies across 
the Dalitz plot, and for this reason we prefer 
$\,\hat\epsilon\,$  as a more natural measure of direct  $CP$  
violation in this mode.

Similarly, for the magnetic amplitudes, we can define
\begin{equation}  \label{eta+-gm}
\eta_{+-\gamma}^{\rm M}  \;\equiv\;  
{ M \bigl( K_{\rm S}^{} \bigr) \over 
 M \bigl( K_{\rm L}^{} \bigr) }  
\;=\;  \epsilon + \epsilon_{+-\gamma}^{\rm M}     \;,  
\end{equation}
where in this case the direct-$CP$-violating parameter is  
\begin{equation}   \label{e+-gm}  
\epsilon_{+-\gamma}^{\rm M}  \;=\;   
{\rm i} \left(   
{ {\rm Im}\, \xi_{\rm M}^{} \over {\rm Re}\, \xi_{\rm M}^{} }
- { {\rm Im}\, A_0^{} \over {\rm Re}\, A_0^{}} 
\right)  
\end{equation}
and we have again dropped terms quadratic in weak phases.  
It appears that it is not possible 
to measure this parameter in an experiment that does not detect 
the photon polarization, as we will see in the next two sections.

It was argued in the past~\cite{DonHV,LinVal} that  
$\tilde{\epsilon}_{+-\gamma}'$  could be several times larger 
than  $\epsilon'$  because it predominantly arises from 
the interference between  $\,\Delta I=1/2\,$  components of 
the IB and DE  terms and, therefore, it is not suppressed by 
the factor 
$\,\omega={\rm Re}\, A_2^{}/{\rm Re}\, A_0^{}.\,$   
Rather, since the IB and DE contributions are generated at  
orders~$p^2$  and~$p^4$,  respectively,  in chiral   
perturbation theory,  $\tilde{\epsilon}_{+-\gamma}'$  
has a suppression factor of  
$\,p^2/\Lambda^2\sim m_K^2/\Lambda^2,\,$   
where  $\,\Lambda\sim 1\,\rm GeV.\,$   
With the additional assumption that the weak phases in  
$\tilde{\epsilon}_{+-\gamma}'$  and  $\epsilon'$  are comparable, 
the enhancement was expected to be\footnote{The original 
argument was done in terms of  $\tilde{\epsilon}_{+-\gamma}'$  
which is related to $\hat\epsilon$  by Eq.~(\ref{e_ppg}). 
The kinematic dependence of  $\tilde{\epsilon}_{+-\gamma}'$  
is such, however, that a considerable suppression results when 
it is integrated over phase space, as we will see in the next 
section.} 
$\,\hat\epsilon/\epsilon'\sim
\bigl( m_K^2/\Lambda^2 \bigr) /\omega\sim 5.\,$ 
Using Eqs.~(\ref{e,e'})  and~(\ref{he}),  we find
\begin{eqnarray}   
\left| {\hat\epsilon\over\epsilon'} \right|  \sim
{ |G_8^{}|\, f_{\!\pi}^2 m_K^{}\, |F_{\rm E}^{}|  \over  
 2\sqrt{2}\, \omega\,  {\rm Re}\, A_0^{} }     
\;\sim\;  
{f_{\!\pi}^{}\, |F_{\rm E}^{}|\over 4 \omega\, m_K^{}}   \;,         
\end{eqnarray}   
From this result, one can see that if  $F_{\rm E}^{}$  is of 
order one as we have speculated, then  
$\,\hat\epsilon \sim \epsilon'\,$  
and there is no enhancement in this reaction.  
Equivalently, for the original estimate  
$\,\hat\epsilon \sim 5 \epsilon'\,$  
to be true, one would need  $\,F_{\rm E}^{} \sim 5\,$.   
It is really impossible to distinguish between these two 
scenarios solely on the grounds of dimensional analysis.    
As we will see in Section~\ref{SM}, an estimate of  
$F_{\rm E}^{}$ based on factorization of the leading 
current-current operator in the effective weak Hamiltonian 
supplemented with vector-meson saturation of the $p^4$ strong 
counterterms suggests that  
$\,F_{\rm E}^{} \sim 1.7\,$  and  
$\hat\epsilon$ is not much larger than 
$\epsilon'$ within the standard model.

A limit for  ${\rm Re}\, F_{\rm E}^{}$  can be obtained from 
the measured  $\,K_{\rm S}^{}\rightarrow\pi^+\pi^-\gamma\,$  
rate, as a nonzero  ${\rm Re}\, F_{\rm E}^{}$  would 
give rise to a difference between the measured rate and 
the calculated IB rate~\cite{SehWan}.  
Including the DE electric-amplitude from Eqs.~(\ref{DEE}) 
and~(\ref{xi_ES}),   
and neglecting the magnetic contribution, we derive\footnote{
In the first term (IB only) we have incorporated 
the complete  $\,K_{\rm S}^{}\rightarrow\pi^+\pi^-\,$  
amplitude as extracted from data~\cite{pdb},  
and in the second term we have used an energy-dependent   
$\,\delta_0^0-\delta_1^1\,$  from  Ref.~\cite{delta}.}   
\begin{eqnarray}   \label{BR,K_S}   
{\rm BR} \bigl( K_{\rm S}^{}\rightarrow\pi^+\pi^-\gamma \bigr)    
\;\simeq\;  
1.75\times 10^{-3} + 1.07\times 10^{-5}\; {\rm Re}\, F_{\rm E}^{} 
+ 2.81\times10^{-8}\; \bigl( {\rm Re}\, F_{\rm E}^{} \bigr) ^2    
\end{eqnarray}   
for  $\,E_\gamma^*>50\,{\rm MeV}.\,$        
Since measurements~\cite{deff1}  give  
$\,{\rm BR} \bigl( K_{\rm S}^{}\rightarrow\pi^+\pi^-\gamma,
E_\gamma^*>50\,{\rm MeV} \bigr)    
=(1.76\pm0.06)\times 10^{-3},\,$  
we extract   
$\,{\rm Re}\, F_{\rm E}^{}\simeq 1$  or  $-380\,$   
using the central value.   
Dropping the second solution, which is unnaturally large, 
we obtain
\begin{eqnarray}   \label{ReF,limit}   
{\rm Re}\, F_{\rm E}^{}  \;=\;  1\pm 6   \;.  
\end{eqnarray}   
If the PDG's number~\cite{pdb}  
$\,{\rm BR} \bigl( K_{\rm S}^{}\rightarrow\pi^+\pi^-\gamma,
E_\gamma^*>50\,{\rm MeV} \bigr)    
=(1.78\pm0.05)\times 10^{-3}\,$  
is used instead, the result is similar,   
$\, {\rm Re}\, F_{\rm E}^{}=3\pm 5.\,$  
Therefore, experimentally the question of the natural size 
of  $\hat\epsilon/\epsilon'$  has not been resolved, and 
a value  $\,\hat\epsilon/\epsilon' \sim 5\,$  as in 
the original dimensional analysis estimate is possible.

\section{Interference between  
$\,K_{\rm L,S}^{}\rightarrow\pi^+\pi^-\gamma\,$  
amplitudes\label{int}}

The parameter  $\hat\epsilon$  may be measured  
in an experiment studying the interference between   
the amplitudes for  $K_{\rm L}^{}$  and  $K_{\rm S}^{}$  
decaying into  $\pi^+\pi^-\gamma$.    
Such an experiment typically~\cite{eta+-g1,eta+-g2} employs 
two~$K_{\rm L}^{}$  beams,  one of which is passed through 
a ``regenerator'', which coherently converts 
some~$K_{\rm L}^{}$  to~$K_{\rm S}^{}$.   
It follows that the initial kaon state can be expressed as 
a coherent mixture  
\begin{eqnarray}   
\bigl| K_{\rm L}^{} \bigr\rangle 
+ \rho\, \bigl| K_{\rm S}^{} \bigr\rangle   \;,  
\end{eqnarray}   
up to a normalization constant, where  
$\,\rho\equiv|\rho|\, {\rm e}^{{\rm i}\phi_\rho^{}}\,$  is 
the regeneration parameter.  
Then the number of decays into  $\pi^+\pi^-\gamma$  
per unit proper time  $\tau$  is given by~\cite{SehWol}
\begin{eqnarray}   \label{G(t)}  
{dN\over d\tau}  \; \propto \;  
%{N_{\rm S}^{}\over|\rho|^2} 
\left\{ 
|\rho|^2\, \Gamma_{K_{\rm S}^{}\rightarrow\pi\pi\gamma}^{}\,  
{\rm e}^{-\tau/\tau_{\rm S}^{}}  
+ \Gamma_{K_{\rm L}^{}\rightarrow\pi\pi\gamma}^{}\,  
 {\rm e}^{-\tau/\tau_{\rm L}^{}}   
+ 2\, {\rm Re} \left[ \rho\, \gamma_{\rm LS}^{*}\, 
                     {\rm e}^{{\rm i}\, \Delta m\, \tau} \right] \, 
{\rm e}^{-(1/\tau_{\rm L}^{}+1/\tau_{\rm S}^{})\tau/2}   
\right\}   \;,  
\end{eqnarray}   
where  
% $N_{\rm S}^{}$  is the number of  $K_{\rm S}^{}$  regenerated,   
$\tau_{\rm L}^{}$  $(\tau_{\rm S}^{})$  is the  $K_{\rm L}^{}$  
$(K_{\rm S}^{})$  lifetime,  $\Delta m$  is  
the $K_{\rm L}^{}$-$K_{\rm S}^{}$  mass difference, 
$\Gamma_{K\rightarrow\pi\pi\gamma}^{}$  is the partial width of 
$\,K\rightarrow\pi^+\pi^-\gamma,\,$  
and  $\gamma_{\rm LS}^{}$  is an integral containing 
the interference between the  $K_{\rm L}^{}$  and  $K_{\rm S}^{}$  
amplitudes.  It is important to keep in mind that the quantities 
$\Gamma_{K\rightarrow\pi\pi\gamma}^{}$  and  $\gamma_{\rm LS}^{}$ 
are {\it not} integrated over all phase space. They depend implicitly 
on the cuts that define the region of phase space under study, 
and in the following discussion it is implied that all the terms 
in Eq.~(\ref{G(t)}) are subject to the {\it same} set of 
kinematic cuts.
  
We now proceed to examine the three terms in Eq.~(\ref{G(t)}) 
in detail. 
To this aim, we will make use of the following relations:
\begin{equation} \label{ratioib}
\Gamma_{K_{\rm L}^{}\rightarrow\pi^+\pi^-\gamma}^{\rm IB}  
\;=\; 
|\eta_{+-}^{}|^2\,  
\Gamma_{K_{\rm S}^{}\rightarrow\pi^+\pi^-\gamma}^{\rm IB}   \;, 
\end{equation}   
which follows from Eq.~(\ref{IB,K_S,K_L});
the ratio~\cite{deff2}  
\begin{eqnarray}   \label{rff}
f  \;\equiv\;  
{ \Gamma_{K_{\rm L}^{}\rightarrow\pi^+\pi^-\gamma}^{\rm M}  \over  
 \Gamma_{K_{\rm L}^{}\rightarrow\pi^+\pi^-\gamma}^{} }  
\;\equiv\;  {r\over 1+r}   
\;\simeq\;  0.685  \;,  
\end{eqnarray}   
which is determined experimentally from a fit to the decay 
spectrum that {\bf assumes} that the interference between the IB and 
an E1-DE is negligible; and Eqs.~(\ref{eta+-g}),~(\ref{eta+-gm}). 
Consequently,  
\begin{eqnarray}   \label{r}
r  \;=\;  
{ \Gamma_{K_{\rm L}^{}\rightarrow\pi^+\pi^-\gamma}^{\rm M}  \over  
 \Gamma_{K_{\rm L}^{}\rightarrow\pi^+\pi^-\gamma}^{\rm E} }   
\;\simeq\;  2.16   \;.     
\end{eqnarray}   

If the photon polarization is not observed, there is no 
interference between the electric and magnetic amplitudes. 
In this case the  
$\,K_{\rm S}^{}\rightarrow\pi^+\pi^-\gamma\,$ 
rate can be decomposed into the sum of electric and magnetic 
rates 
$$\Gamma_{K_{\rm S}^{}\rightarrow\pi^+\pi^-\gamma}^{\rm E} + 
\Gamma_{K_{\rm S}^{}\rightarrow\pi^+\pi^-\gamma}^{\rm M}\;.$$  
Since the second term is $CP$-violating, it is convenient to 
rewrite it in terms of Eq.~(\ref{eta+-gm})  
as\footnote{We have here assumed  $\eta_{+-\gamma}^{\rm M}$  
to be a constant, which is appropriate for our purposes.}   
\begin{eqnarray}   \label{shortrate}
\Gamma_{K_{\rm S}^{}\rightarrow\pi^+\pi^-\gamma}^{} 
&=&  
\Gamma_{K_{\rm S}^{}\rightarrow\pi^+\pi^-\gamma}^{\rm E} +
\bigl| \eta_{+-\gamma}^{\rm M} \bigr| ^2\,  
\Gamma_{K_{\rm L}^{}\rightarrow\pi^+\pi^-\gamma}^{\rm M}  
\nonumber \\  
&\simeq&  
\left( 
1 + \bigl| \eta_{+-}^{} \bigr| ^2\,  
\bigl| \eta_{+-\gamma}^{\rm M} \bigr| ^2\, r   
\right)  
\Gamma_{K_{\rm S}^{}\rightarrow\pi^+\pi^-\gamma}^{\rm E}   
\;\simeq\;   
\Gamma_{K_{\rm S}^{}\rightarrow\pi^+\pi^-\gamma}^{\rm E}   \;.  
\end{eqnarray}   

The interference term in Eq.~(\ref{G(t)}) can be written in the 
kaon rest-frame as  
\begin{eqnarray}   
\gamma_{\rm LS}^{}  \;=\;  
\int{\rm d[PS]}\; \Bigl\{ 
\left[ E_{\rm IB}^{} \bigl( K_{\rm L}^{} \bigr) ^{}   
      + E_{\rm DE}^{} \bigl( K_{\rm L}^{} \bigr) \right] 
\left[ E_{\rm IB}^{*} \bigl( K_{\rm S}^{} \bigr)    
      + E_{\rm DE}^{*} \bigl( K_{\rm S}^{} \bigr) \right] _{}^{}
+ M(K_{\rm L}^{})\, M^*(K_{\rm S}^{}) 
\Bigr\}   \;,  
\end{eqnarray}   
where  
\begin{eqnarray}   
{\rm d[PS]}  \;\equiv\;  
{\rm d\,cos}\theta\, {\rm d}E_\gamma^*\; 
{ \bigl( \beta E_\gamma^* \bigr) ^3 {\rm sin}^2\theta  \over 
 32\pi^3\, m_K^3 }  
\left( 1-{2E_\gamma^*\over m_K^{}} \right)   \;,  
\end{eqnarray}   
with  $\theta$  being the angle between  
$\mbox{\boldmath$p$}_+^{}$  and  $\mbox{\boldmath$q$}$  in 
the  $\pi\pi$  rest-frame,  and  
$\,\beta=\sqrt{1-4m_{\pi}^2/( m_K^2-2E_\gamma^*m_K^{} ) }.\,$  
We remark that  $\,\rm d[PS]\,$  contains not only 
the phase-space factor, but also factors that resulted from 
summing over the photon polarizations and from contraction of 
tensor forms.  
Making use of the definitions in  
Eqs.~(\ref{eta+-g}), (\ref{eta+-gm}),~(\ref{rff}) and 
defining\footnote{Notice that {\it both} 
$\Gamma_{K_{\rm S}\rightarrow\pi^+\pi^-\gamma}^{}$  
in the denominator and  the integral in the numerator of  
Eq.~(\ref{te+-g'})  depend on the  $E_\gamma^*$  cut.}
\begin{eqnarray}   \label{te+-g'}  
\epsilon_{+-\gamma}'  &\equiv&  
{1\over\Gamma_{K_{\rm S}^{}\rightarrow\pi^+\pi^-\gamma}^{}}  
\int{\rm d[PS]}\; \tilde{\epsilon}_{+-\gamma}' \,      
\left| E_{\rm IB}^{} \bigl( K_{\rm S}^{} \bigr) 
      + E_{\rm DE}^{} \bigl( K_{\rm S}^{} \bigr) \right| ^2   \;,  
\end{eqnarray}   
one finds that   
\begin{eqnarray}   \label{g_LS}   
\gamma_{\rm LS}^{}  \;=\;  
\left\{  
\eta_{+-}^{} + \epsilon_{+-\gamma}' 
\,+\, 
\eta_{+-\gamma}^{\rm M*}\, r\, \left[ 
\bigl| \eta_{+-}^{} \bigl| ^2 
\,+\,  
2\, {\rm Re} \bigl( \eta_{+-}^*\, \epsilon_{+-\gamma}' \bigr) 
\right] 
\right\} 
\Gamma_{K_{\rm S}^{}\rightarrow\pi^+\pi^-\gamma}^{}   \;,    
\end{eqnarray}   
having dropped terms suppressed by additional powers of  
$\epsilon$  or  $\hat\epsilon$.   

Turning now to the second term in Eq.~(\ref{G(t)}), we have  
\begin{eqnarray}   
\Gamma_{K_{\rm L}^{}\rightarrow\pi^+\pi^-\gamma}^{}  \;=\;   
\int{\rm d[PS]}\; \left( 
\left| E_{\rm IB}^{} \bigl( K_{\rm L}^{} \bigr)    
      + E_{\rm DE}^{} \bigl( K_{\rm L}^{} \bigr) \right| ^2 
+ \left| M \bigl( K_{\rm L}^{} \bigr) \right| ^2 
\right)   \;,  
\end{eqnarray}   
where  the photon polarizations have been summed over.  
Then, from  Eqs.~(\ref{eta+-g}), (\ref{te+-g'}),  
and~(\ref{rff}),  neglecting terms of orders  
$\hat\epsilon^2$  and  $\epsilon^6$,  we find   
\begin{eqnarray}   
\Gamma_{K_{\rm L}^{}\rightarrow\pi^+\pi^-\gamma}^{}  \;=\;   
(1+r) \left[ 
\bigl| \eta_{+-}^{} \bigr| ^2 
+ 2\, {\rm Re}\left( \eta_{+-}^{*}\epsilon'_{+-\gamma}\right) \right]
\Gamma_{K_{\rm S}^{}\rightarrow\pi^+\pi^-\gamma}^{}  \;. 
\end{eqnarray}   

Collecting all these results, we can now rewrite  
Eq.~(\ref{G(t)})  as  
\begin{eqnarray}   \label{rate,t}
{dN\over d\tau}  \; \propto \;  
%{ N_{\rm S}^{}\,  \over  |\rho|^2 }
\Gamma_{K_{\rm S}^{}\rightarrow\pi^+\pi^-\gamma}^{}  
\begin{array}[t]{l}   \displaystyle     
\! \left\{  
|\rho|^2\, {\rm e}^{-\tau/\tau_{\rm S}^{}}  
\,+\,  
\left[ 
\bigl| \eta_{+-}^{} \bigr| ^2   
+ 2\, {\rm Re} \left( \eta_{+-}^* \epsilon'_{+-\gamma} \right) 
\right] (1+r)\, {\rm e}^{-\tau/\tau_{\rm L}^{}}   
\right. 
\vspace{2ex} \\   \displaystyle     
\left. 
+\;  
2\, \left| 
\eta_{+-}^{} 
+ \epsilon'_{+-\gamma} \right| \, |\rho|\;  
{\rm cos} \left( \Delta m\,\tau+\phi_\rho^{} 
                - \phi_\eta^{} \right) \, 
{\rm e}^{-\tau(1/\tau_{\rm L}^{}+1/\tau_{\rm S}^{})/2}   
\right\}   \;,        
\end{array}   
\end{eqnarray}   
where  ${\phi}_\eta^{}$  is the phase of 
$\,\bigl(\eta_{+-}^{} + \epsilon_{+-\gamma}'\bigr).\,$  
This rate equation can be used to extract  
$\hat{\epsilon}$  from measurements.  
The most recent experimental study~\cite{eta+-g2}  
on these decays starts from the definition  
\begin{eqnarray}   \label{rate,x}
{dN\over d\tau}  \;=\;  
{ N_{\rm S}^{}\, \Gamma_{K_{\rm S}^{}\rightarrow\pi^+\pi^-\gamma}^{} 
 \over  |\rho|^2 } 
\begin{array}[t]{l}   \displaystyle     
\! \left\{  
|\rho|^2\, {\rm e}^{-\tau/\tau_{\rm S}^{}}  
\,+\,  
\bigl| \eta_{+-\gamma}^{} \bigr| ^2\, (1+r)\,  
{\rm e}^{-\tau/\tau_{\rm L}^{}}   
\right. 
\vspace{2ex} \\   \displaystyle     
\left. 
+\;  
2\, \bigl| \eta_{+-\gamma}^{} \bigr| \, |\rho|\;  
{\rm cos} \left( \Delta m\,\tau+\phi_\rho^{}-\phi_\eta^{} \right) \, 
{\rm e}^{-\tau(1/\tau_{\rm L}^{}+1/\tau_{\rm S}^{})/2}   
\right\}   \;,           
\end{array}   
\end{eqnarray}   
where  $N_{\rm S}^{}$  is the number of  $K_{\rm S}^{}$  regenerated.    
Comparing Eqs.~(\ref{rate,t})  and~(\ref{rate,x}), we see that
\begin{equation}
\eta_{+-\gamma}^{}  \;\equiv\;  
\eta_{+-}^{} + \epsilon'_{+-\gamma}
\end{equation}
if terms of order $(\hat\epsilon/\epsilon)^2$ are neglected.

With a minimum photon energy, $\,E_\gamma^*>20\,{\rm MeV},\,$  
we integrate over phase space ($\,-1<\cos\theta<1\,$  and  
$\,20\,{\rm MeV}<E_\gamma^*<m_K^{}/2-2m_\pi^2/m_K^{} ,\,$) 
and use 
$\,E \bigl( K_{\rm S}^{} \bigr) \sim 
E_{\rm IB}^{} \bigl( K_{\rm S}^{} \bigr),\,$  
to find  
\begin{eqnarray}   \label{te+-g''}  
\epsilon_{+-\gamma}'  \;\simeq\;  
0.041\; \hat\epsilon\, {\rm e}^{{\rm i}\delta_\epsilon^{}}   \;,  
\end{eqnarray}   
where we have also used    
$\,{\rm BR} \bigl( K_{\rm S}^{}\rightarrow\pi^+\pi^-\gamma,   
E_\gamma^*>20\,{\rm MeV} \bigr)   
=4.87\times10^{-3}\,$  
from  Ref.~\cite{deff1}  and  $\,e^2/(4\pi)=1/137,\,$  
and  assumed  $\delta_\epsilon^{}$  to be constant.  
We have checked that a phase which varies with energy 
would not alter this result in a significant way.  
With an energy-dependent formula for  
$\,\delta_0^0-\delta_1^1\,$  from  Ref.~\cite{delta},  
we get  
$\,\epsilon_{+-\gamma}'\simeq  
(0.014+0.039\;{\rm i})\, \hat\epsilon,\,$  
which is not different from  Eq.~(\ref{te+-g''})  if 
$\,\delta_0^0-\delta_1^1\,$  takes its average value  
of~$\,17.4^{\rm o}.\,$

We see that the observable  $\epsilon_{+-\gamma}'$  is 
suppressed by a factor of  $\,0.041\,$  with respect to  
$\hat\epsilon$.  
This factor can be understood as the ratio of 
$\,\int\mbox{d(phase space)}/(z^2-\nu^2)\,$  
to  $\,\int\mbox{d(phase space)}/(z^2-\nu^2)^2,\,$    
which are the corresponding forms for the interference and 
IB terms in the rate.  
It is interesting to notice that the suppression factor 
decreases as the cut in  $E_\gamma^*$  is increased.  
This is due to the fact that with higher cuts  
the ratio of the IB contribution to that of the electric DE 
becomes smaller.  
For example, we find  
\begin{eqnarray}   \label{te+-g'''}  
\epsilon_{+-\gamma}'  \;\simeq\;   
0.090\; \hat\epsilon\, {\rm e}^{{\rm i}\delta_\epsilon^{}}  
\hspace{1em} \mbox{for  $\,E_\gamma^*>50\,{\rm MeV},\,$}  
\end{eqnarray}   
by means of  
$\,{\rm BR} \bigl( K_{\rm S}^{}\rightarrow\pi^+\pi^-\gamma,
E_\gamma^*>50\,{\rm MeV} \bigr)   
=1.76\times 10^{-3}\,$  
from  Ref.~\cite{deff1}.   
Clearly if the  $E_\gamma^*$  cut is increased much 
further, one has to reconsider the assumption of 
IB~dominance in  the  $K_{\rm S}^{}$  amplitude.   
Experimentally, of course, an increased $E_\gamma^*$ cut 
results in a smaller event sample.
 
The best measurement  $\bigl(E_\gamma^*>20\,{\rm MeV}\bigr)$  
to date~\cite{eta+-g3},  
\begin{eqnarray} 
\left| {\epsilon_{+-\gamma}'\over\epsilon} \right| _{\rm exp}^{} 
\;=\;  0.041\pm0.035   \;,  
\end{eqnarray}   
translates into the one-sigma limit   
\begin{eqnarray}   \label{|he/e|_x}     
\left| {\hat\epsilon\over\epsilon} \right|  \;<\;  1.9   \;.      
\end{eqnarray}   
Then, if we assume that there is no cancellation between 
the two phases in  Eq.~(\ref{he}),  this implies that  
${\rm Im}\, F_{\rm E}^{}$  could have a magnitude as large as  
\begin{eqnarray}   \label{F_E}     
\left| {\rm Im}\, F_{\rm E}^{} \right|  \;\sim\;  0.1   \;.      
\end{eqnarray}   
The KTeV experiment is expected~\cite{eta+-g3} to reduce 
the uncertainties in   
$ \bigl| \epsilon_{+-\gamma}'/\epsilon \bigr| $  
to the  $\,0.4\%\,$  level,  which would improve these bounds 
by a factor of~10.

To end this section we comment on the observability of 
the quantity  $\eta_{+-\gamma}^{\rm M}$  for  $CP$ violation in 
the $K_{\rm S}^{}$  magnetic-amplitude.  
It is evident from Eqs.~(\ref{shortrate})  and~(\ref{g_LS})  
that  $\eta_{+-\gamma}^{\rm M}$  occurs amongst terms 
of order  $\epsilon^3$  or higher.   
Therefore, it does not seem possible to measure   
the direct-$CP$-violation parameter  
$\epsilon_{+-\gamma}^{\rm M}$,  defined in 
Eq.~(\ref{e+-gm}), in this kind of experiment.

\section{Interference between IB and DE in electric 
amplitude\label{IBDEE}}

The direct $CP$  violating parameter $\hat\epsilon$ can also be 
detected in principle by analyzing in detail the decay 
distribution in 
$\,K_{\rm L}^{}\rightarrow\pi^+\pi^-\gamma\,$.  
Let us first consider the photon-energy   
$(E_\gamma^*)$  spectrum used to separate IB from DE contributions. 
The  $E_\gamma^*$  distribution is given by  
\begin{eqnarray}   \label{dG/dE}
{ d\Gamma_{K_{\rm L}^{}\rightarrow\pi^+\pi^-\gamma}^{}   
 \over  d E_\gamma^*}   
\;=\;  
{ \bigl( \beta E_\gamma^* \bigr) ^3\over 32\pi^3\, m_K^3} 
\left( 1-{2E_\gamma^*\over m_K^{}} \right) 
\int_{-1}^1{\rm sin}^2\theta\, {\rm d cos}\theta\, \left( 
\left| E_{\rm IB}^{} \bigl( K_{\rm L}^{} \bigr)    
      + E_{\rm DE}^{} \bigl( K_{\rm L}^{} \bigr) \right| ^2 
+ \left| M \bigl( K_{\rm L}^{} \bigr) \right| ^2 
\right)   \;.   
\end{eqnarray}   
Recent experimental studies~\cite{deff1,deff2} of  
the  $E_\gamma^*$  spectrum assume that the DE process  
is purely magnetic and parameterize it with 
a form-factor-modified M1 contribution. 
By parameterizing the  $CP$-violating  E1  component in the DE electric 
amplitude as in Eq.~(\ref{xEDE}) we can calculate  
its interference with the IB contribution. The presence of this 
term modifies the shape of the $(E_\gamma^*)$  spectrum, and 
this generates a limit on $\hat\epsilon$.
  
We present in  Fig.~\ref{dG/dE,plot}  an example of such 
a deviation.  
For this figure, the magnetic amplitude is parameterized in the 
form used by recent experiments~\cite{deff1,deff2}, 
\begin{eqnarray}   \label{expdef}
\xi_{\rm M}^{\rm L}  \;=\; \left( 
{a_1^{}\over m_\rho^2-m_K^2 + 2E_\gamma^* m_K^{}} 
+ a_2^{} \right) \, {\rm e}^{{\rm i}\delta^1_1}   \;,  
\end{eqnarray}   
where  $m_\rho$  is the  $\rho$-meson mass,   
and  $a_{1,2}^{}$  are parameters obtained from the 
experimental fit.   
The solid curve describes the combination of the IB and   
magnetic DE processes, without any electric DE contribution.   
For this curve, we employ  
$\,a_1^{}/a_2^{}=-0.729\, \rm GeV^2\,$  
from Ref.~\cite{deff2}  and  adjust the value of  $a_{2}^{}$  
in order to get a  
$\,K_{\rm L}^{}\rightarrow\pi^+\pi^-\gamma\,$   
rate that matches the measured one  
$\bigl($thus here  
$\,a_{2}^{}\simeq 3.09 \bigr).$  
The upper dashed-curve includes the IB, the magnetic DE 
with  $a_1^{}/a_2^{}$ as before and 
an IB--E1-DE interference using $\,{\rm Re}\, F_{\rm E}^{}=0\,$  and 
$\,{\rm Im}\, F_{\rm E}^{}=0.5\,$ for illustration. In this case 
we have used $\,a_{2}^{}\simeq 2.40$ to keep the total rate fixed.  
The lower dashed-curve corresponds to the interference between 
the IB and E1 DE only.

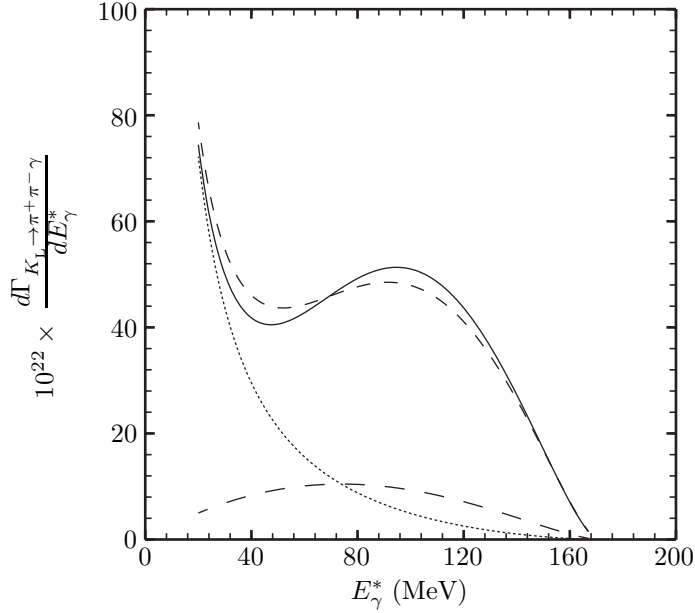
\begin{figure}[ht]         
   \hspace*{\fill}   
\begin{picture}(300,250)(-50,-25)   
% y axis 
\LinAxis(0,0)(0,200)(5,5,-3,0,1) \LinAxis(200,0)(200,200)(5,5,3,0,1) 
% x axis 
\LinAxis(0,0)(200,0)(5,5,3,0,1) \LinAxis(0,200)(200,200)(5,5,-3,0,1)  
% (2 10^22 dWidth/dE, 1000 E) 
%  
% IB & DEM,  fE = sE = 0,  a2 = 7.28109 10^-8   
\Curve{(20., 148.869)(23., 128.523)(26., 113.74)(29., 102.905) 
(32., 94.9945)(35., 89.3218)(38., 85.4031)(41., 82.8816) 
(44., 81.4836)(47., 80.9919)(50., 81.2286)(53., 82.0441) 
(56., 83.3103)(59., 84.915)(62., 86.7589)(65., 88.7533) 
(68., 90.818)(71., 92.8808)(74., 94.876)(77., 96.7442) 
(80., 98.4315)(83., 99.8897)(86., 101.075)(89., 101.95) 
(92., 102.481)(95., 102.639)(98., 102.4)(101., 101.745) 
(104., 100.657)(107., 99.1277)(110., 97.1496)(113., 94.7215) 
(116., 91.8459)(119., 88.5303)(122., 84.7863)(125., 80.6302) 
(128., 76.0832)(131., 71.1712)(134., 65.9254)(137., 60.3824) 
(140., 54.585)(143., 48.5825)(146., 42.4321)(149., 36.2) 
(152., 29.9639)(155., 23.8162)(158., 17.8695)(161., 12.2667) 
(164., 7.20137)(167., 2.96891)} 
%  
% IB + DE,  fE = 0.5,  sE = 0,  a2 = 5.65097 10^-8  
\DashCurve{(20., 157.294)(23., 137.49)(26., 123.049)(29., 112.362) 
(32., 104.417)(35., 98.5387)(38., 94.2581)(41., 91.2338) 
(44., 89.2091)(47., 87.9844)(50., 87.3998)(53., 87.3239) 
(56., 87.6467)(59., 88.2736)(62., 89.1226)(65., 90.1212) 
(68., 91.205)(71., 92.3159)(74., 93.4017)(77., 94.4149) 
(80., 95.3125)(83., 96.0554)(86., 96.6081)(89., 96.9388) 
(92., 97.0188)(95., 96.8224)(98., 96.3273)(101., 95.5141) 
(104., 94.3662)(107., 92.8703)(110., 91.0158)(113., 88.7952) 
(116., 86.2043)(119., 83.2418)(122., 79.9098)(125., 76.2141) 
(128., 72.1641)(131., 67.7731)(134., 63.0592)(137., 58.0454) 
(140., 52.7602)(143., 47.2392)(146., 41.5258)(149., 35.6734) 
(152., 29.7479)(155., 23.8321)(158., 18.0321)(161., 12.4887) 
(164., 7.40074)(167., 3.08137)}{5} 
% 
% IB-DEE interference only,  fE = 0.5,  sE = 0  
\DashCurve{(20., 9.98057)(23., 11.1906)(26., 12.3269)(29., 13.3899) 
(32., 14.38)(35., 15.2978)(38., 16.1436)(41., 16.9181) 
(44., 17.6218)(47., 18.2552)(50., 18.8189)(53., 19.3135) 
(56., 19.7397)(59., 20.0982)(62., 20.3897)(65., 20.6148) 
(68., 20.7745)(71., 20.8696)(74., 20.9009)(77., 20.8694) 
(80., 20.776)(83., 20.6219)(86., 20.4081)(89., 20.1358) 
(92., 19.8063)(95., 19.4209)(98., 18.9811)(101., 18.4883) 
(104., 17.9444)(107., 17.351)(110., 16.7101)(113., 16.0238) 
(116., 15.2943)(119., 14.5242)(122., 13.7161)(125., 12.8729) 
(128., 11.9981)(131., 11.0951)(134., 10.1681)(137., 9.22151) 
(140., 8.26054)(143., 7.29103)(146., 6.31971)(149., 5.35446) 
(152., 4.40464)(155., 3.48159)(158., 2.5995)(161., 1.77682) 
(164., 1.03926)(167., 0.427109)}{7} 
%   
% IB only 
\DashCurve{(20., 144.362)(23., 122.044)(26., 104.899)(29., 91.3219) 
(32., 80.3094)(35., 71.2024)(38., 63.5498)(41., 57.0327) 
(44., 51.4191)(47., 46.5363)(50., 42.2529)(53., 38.4675) 
(56., 35.1002)(59., 32.0875)(62., 29.3783)(65., 26.9306) 
(68., 24.7102)(71., 22.6885)(74., 20.8416)(77., 19.1493) 
(80., 17.5944)(83., 16.1624)(86., 14.8405)(89., 13.618) 
(92., 12.4855)(95., 11.4345)(98., 10.4581)(101., 9.54973) 
(104., 8.70392)(107., 7.91572)(110., 7.18074)(113., 6.4951) 
(116., 5.85535)(119., 5.2584)(122., 4.70152)(125., 4.18227) 
(128., 3.69847)(131., 3.24821)(134., 2.8298)(137., 2.44175) 
(140., 2.08281)(143., 1.75189)(146., 1.44813)(149., 1.1709) 
(152., 0.919781)(155., 0.69468)(158., 0.495876)(161., 0.324213) 
(164., 0.181478)(167., 0.0714074)}{1} 
\footnotesize  
% y-axis label
\rText(-40,100)[][l]{$10^{22}\times\; 
{\textstyle d\Gamma_{K_{\rm L}^{}\rightarrow\pi^+\pi^-\gamma}^{}   
 \over  \textstyle  d E_\gamma^* }   
$}   
% x-axis label 
\Text(100,-15)[t]{$E_\gamma^*\;(\rm MeV)$}   
% y points   
\Text(-3,200)[r]{$100$} \Text(-3,160)[r]{$80$} 
\Text(-3,120)[r]{$60$} \Text(-3,80)[r]{$40$} 
\Text(-3,40)[r]{$20$} \Text(-3,0)[r]{$0$}  
% x-points  
\Text(0,-3)[t]{$0$} \Text(40,-3)[t]{$40$} \Text(80,-3)[t]{$80$} 
\Text(120,-3)[t]{$120$} \Text(160,-3)[t]{$160$} 
\Text(200,-3)[t]{$200$} 
\end{picture} 
   \hspace*{\fill} 
\caption{\label{dG/dE,plot}%  
Photon-energy $(E_\gamma^*)$  distribution in 
$\,K_{\rm L}^{}\rightarrow\pi^+\pi^-\gamma.\,$  
The solid curve represents the sum of the 
inner bremsstrahlung and the magnetic direct-emission 
contributions. The dotted curve is the IB alone.  
The upper dashed-curve corresponds to IB plus M1-DE contributions 
as in the solid curve, plus an IB--E1-DE interference with 
$\,F_{\rm E}^{}=0.5\rm i,\,$  as explained in the text. 
The lower dashed-curve shows the interference between 
the IB and E1 DE only.}  
\end{figure}             

From  Eq.~(\ref{dG/dE}),  we find the interference between 
the IB and DE components of the electric amplitude to be  
\begin{eqnarray}   
\Gamma_{K_{\rm L}^{}\rightarrow\pi^+\pi^-\gamma}^{\rm E,int}   
\;=\;   
\int{\rm d[PS]}\; 2\, {\rm Re} \left[   
E_{\rm IB}^{*} \bigl( K_{\rm L}^{} \bigr)    
E_{\rm DE}^{} \bigl( K_{\rm L}^{} \bigr)    
\right]   \;,  
\end{eqnarray}   
with the integration ranges being 
$\,-1<\cos\theta<1\,$  and  
$\,20\,{\rm MeV}<E_\gamma^*<m_K^{}/2-2m_\pi^2/m_K^{}.\,$ 
Using  
\begin{eqnarray} 
\Gamma_{K_{\rm L}^{}\rightarrow\pi^+\pi^-\gamma}^{} 
\bigl( E_\gamma^*>20\,{\rm MeV} \bigr)  
\;=\;  5.86\times 10^{-19}\; {\rm MeV}   
\end{eqnarray}   
from  Ref.~\cite{pdb},   
Eqs.~(\ref{IB,K_S,K_L})  and~(\ref{xi_EL}), as well as 
$\,\delta_0^0-\delta_1^1=17.4^{\rm o}\,$  
in  $\delta_\epsilon^{}$,  we obtain   
\begin{eqnarray}   \label{int/tot}  
{ \Gamma_{K_{\rm L}^{}\rightarrow\pi^+\pi^-\gamma}^{\rm E,int}   
 \over  \Gamma_{K_{\rm L}^{}\rightarrow\pi^+\pi^-\gamma}^{} }  
&\simeq&  
5.0\times10^3\; \bigl| \eta_{+-}^{} \bigr| \, \hat\epsilon\, 
{\rm cos}(\delta_\epsilon^{}-\phi_{+-}^{}) 
+ 1.8\times10^2\; \bigl| \epsilon\, \eta_{+-}^{} \bigr| \, 
 {\rm Re}\, F_{\rm E}^{}\, {\rm cos}(\delta_\epsilon^{}-\pi/2)   
\nonumber \\ 
&\simeq&  
0.023\; {\hat\epsilon\over|\epsilon|}  
+ 8.9\times10^{-4}\; {\rm Re}\, F_{\rm E}^{}   \;.  
\end{eqnarray}   
An analysis of KTeV data 
with the same  $E_\gamma^*$  cut, gives~\cite{deff3},    
\begin{eqnarray} \label{ktevbound}   
{ \Gamma_{K_{\rm L}^{}\rightarrow\pi^+\pi^-\gamma}^{\rm E,int}   
 \over  \Gamma_{K_{\rm L}^{}\rightarrow\pi^+\pi^-\gamma}^{} }  
\;<\;  0.30\; \mbox{ (90$\%$ C.L.)}   \;.       
\end{eqnarray}   
Taken literally, this bound implies that,
\begin{eqnarray}   
\left| {\hat\epsilon\over\epsilon} 
      + 0.04\; {\rm Re}\, F_{\rm E}^{} \right|  
\;<\;  13   \;,  
\end{eqnarray}   
and assuming that the two terms do not cancel,
\begin{eqnarray}   \label{ImF}  
\bigl| {\rm Im}\, F_{\rm E}^{} \bigr|  
\;\raisebox{-0.4ex}{$\,\stackrel{\textstyle<}{\textstyle\sim}\,$}\;  
1  \;,  
\hspace{3em}  
\bigl| {\rm Re}\, F_{\rm E}^{} \bigr|  
\;\raisebox{-0.4ex}{$\,\stackrel{\textstyle<}{\textstyle\sim}\,$}\;  
350   \;.   
\end{eqnarray}
These limits are much weaker than Eqs.~(\ref{|he/e|_x}), 
(\ref{F_E}),  and~(\ref{ReF,limit}). However, it is 
important to notice that the bound, Eq.~(\ref{ktevbound}), is not 
particularly strong. In fact, it is amazing that this 
interference between the IB and E1-DE contributions, 
which is expected to be small, could account for as much 
as one third of the experimental rate. This is due in part 
to the fact that an analysis of the full $E_\gamma^*$ distribution 
is not the optimal observable to isolate this term as we 
can see in Fig.~\ref{dG/dE,plot}.

It may be possible to improve the bound, Eq.~(\ref{ktevbound}) 
by restricting the analysis to the region of the 
$E_\gamma^*$ distribution where this term is most important. 
A glance at the dashed curves in  Fig.~\ref{dG/dE,plot}  
suggests that the cuts 
$\,50\,{\rm MeV}<E_\gamma^*<90\,{\rm MeV},\,$ 
for example, would improve the bound. With these cuts, we find 
\begin{eqnarray}   
{ \Gamma_{K_{\rm L}^{}\rightarrow\pi^+\pi^-\gamma}^{\rm E,int}   
 \over  \Gamma_{K_{\rm L}^{}\rightarrow\pi^+\pi^-\gamma}^{} }  
&\simeq&  
0.029\; {\hat\epsilon\over|\epsilon| }  
+ 1.1\times10^{-3}\; {\rm Re}\, F_{\rm E}^{}   \;,  
\end{eqnarray}   
where we used $\,\delta_0^0-\delta_1^1=21.3^{\rm o}\,$  
(the average phase difference in this region) and  
\begin{eqnarray}   
\Gamma_{K_{\rm L}^{}\rightarrow\pi^+\pi^-\gamma}^{} 
\bigl( 50\,{\rm MeV}<E_\gamma^*<90\,{\rm MeV} \bigr)  
\;\simeq\;  1.84\times10^{-19}\; {\rm MeV}   \;,  
\end{eqnarray}   
obtained by including only the IB and magnetic contributions 
(since these should dominate the rate).\footnote{
For the magnetic part,  $\,a_1^{}/a_2^{}=-0.729\, \rm GeV^2\,$  
and  $\,a_{2}^{}\simeq 3.09\,$  have been used.}  
These numbers suggest that this simple set of cuts could 
improve the bound on $\hat\epsilon$ by about 30\%. However, 
this procedure depends on the assumed 
$E_\gamma^*$ dependence of $\xi_{\rm M}^{\rm L}$, Eq.~(\ref{expdef}),
and of $F_{\rm E}$ (independent of $E_\gamma^*$).   

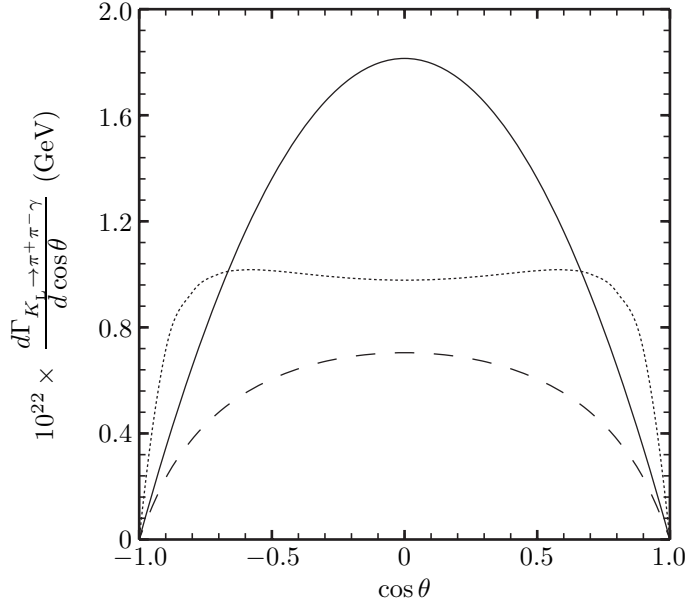
\begin{figure}[hbt]         
   \hspace*{\fill}   
\begin{picture}(300,250)(-150,-25)   
% y axis 
\LinAxis(-100,0)(-100,200)(5,5,-3,0,1) \LinAxis(100,0)(100,200)(5,5,3,0,1) 
% x axis 
\LinAxis(-100,0)(100,0)(4,5,3,0,1) \LinAxis(-100,200)(100,200)(4,5,-3,0,1)  
% (100 10^22 dWidth/dE, 100 Costheta) 
%  
% IB only 
\DashCurve{(-100., 0)(-90., 71.5385)(-80., 93.2603)(-70., 100.115) 
(-60., 101.715)(-50., 101.376)(-40., 100.408)(-30., 99.3744) 
(-20., 98.5289)(-10., 97.9883)(0, 97.8034)(10., 97.9883) 
(20., 98.5289)(30., 99.3744)(40., 100.408)(50., 101.376) 
(60., 101.715)(70., 100.115)(80., 93.2603)(90., 71.5385)(100., 0)}{1} 
%  
% DEM only  a1/a2=-0.729   a2=2.40   
\Curve{(-100., 0)(-90., 34.4715)(-80., 65.3144)(-70., 92.5288) 
(-60., 116.115)(-50., 136.072)(-40., 152.4)(-30., 165.1) 
(-20., 174.172)(-10., 179.615)(0, 181.429)(10., 179.615) 
(20., 174.172)(30., 165.1)(40., 152.4)(50., 136.072) 
(60., 116.115)(70., 92.5288)(80., 65.3144)(90., 34.4715)(100., 0)}
%  
% IB-DEE interference,  fE = 0.5,  sE = 0  
\DashCurve{(-100., 0)(-90., 23.1551)(-80., 37.8707)(-70., 47.9812) 
(-60., 55.2272)(-50., 60.5231)(-40., 64.3947)(-30., 67.1613) 
(-20., 69.021)(-10., 70.0938)(0, 70.4445)(10., 70.0938) 
(20., 69.021)(30., 67.1613)(40., 64.3947)(50., 60.5231) 
(60., 55.2272)(70., 47.9812)(80., 37.8707)(90., 23.1551)(100., 0)}{7}  
\footnotesize  
% y-axis label 
\rText(-140,100)[][l]{$10^{22}\times\; 
{\textstyle d\Gamma_{K_{\rm L}^{}\rightarrow\pi^+\pi^-\gamma}^{}   
 \over \textstyle d \cos\theta}    
\hspace{1ex}(\rm GeV)$}   
% x-axis label 
\Text(0,-15)[t]{$\cos\theta$}   
% y points   
\Text(-103,200)[r]{$2.0$} \Text(-103,160)[r]{$1.6$} 
\Text(-103,120)[r]{$1.2$} \Text(-103,80)[r]{$0.8$} 
\Text(-103,40)[r]{$0.4$} \Text(-103,0)[r]{$0$}  
% x-points  
\Text(-100,-3)[t]{$-1.0$} \Text(-50,-3)[t]{$-0.5$} \Text(0,-3)[t]{$0$} 
\Text(50,-3)[t]{$0.5$} \Text(100,-3)[t]{$1.0$} 
\end{picture} 
   \hspace*{\fill} 
\caption{\label{dG/dc}%
Distributions of various components of the 
$\,K_{\rm L}^{}\rightarrow\pi^+\pi^-\gamma\,$ rate 
in terms of  $\cos\theta$,    
where  $\theta$  is the angle between the  $\pi^+$  and  
$\gamma$  momenta in the  $\pi\pi$  rest-frame.  
The solid curve represents only the magnetic contribution. 
The dotted curve is the IB alone.  
The dashed-curve corresponds to the IB--E1-DE interference with 
$\,F_{\rm E}^{}=0.5\rm i.\,$}  
\end{figure}

A more detailed analysis of the decay distribution is desirable 
in order to separate all the contributions. Unfortunately, we find 
that the decay distributions for the M1 term and for the 
IB-E1 interference term are remarkably similar when the photon 
polarization is not observed. We illustrate this by presenting 
a $\cos\theta$ distribution in Fig.~\ref{dG/dc}. We can see in this 
figure that the shapes of the M1 distribution (solid line), and 
IB-E1 interference term (dashed line) are very similar. The only 
significant difference between the two is that the interference 
term could be negative, depending on the phase of $F_{\rm E}$. 

\section{Standard-model contribution\label{SM}}
   
In this section we review the estimate of  $F_{\rm E}^{}$ 
that exists in the literature and we apply it to an 
estimate of $\hat\epsilon$ within the standard model.  

The~SM  contribution to  the~DE  electric-amplitude in  
$\,K_{\rm L}^{}\rightarrow\pi^+\pi^-\gamma\,$    
is generated at short distance by both four-quark and 
two-quark  $(s\rightarrow d\gamma,d g)$  operators with known 
coefficients. The matrix elements of these operators 
arise at~${\cal O}(p^4)$ in  chiral perturbation theory. 
They are presumably dominated by the chiral 
realizations of the  four-quark operators that transform as  
$\bigl( 8_{\rm L}^{},1_{\rm R}^{} \bigr)$  
under chiral rotations, as well as  by~${\cal O}(p^4)$  
loop  diagrams~\cite{DamIsi2,DamMS}.   
The contribution from the  $\,s\rightarrow d\gamma,d g\,$  
operators in the SM starts at order~$p^6$, and since the 
coefficients of these two-quark operators are quite small, 
it is appropriate to neglect them in the standard model estimate. 

The  ${\cal O}(p^4)$  weak chiral Lagrangian 
transforming as  $ \bigl( 8_{\rm L}^{},1_{\rm R}^{} \bigr) $    
is given by~\cite{DamMS,EckPR}     
\begin{eqnarray}     
{\cal L}_{\rm w}^{(4)}  \;=\;  
G_8^{} f_\pi^2 \sum_{i=14}^{17} N_i^{} W_i^{}   
\;+\;  {\rm H.c.}   \;,    
\end{eqnarray}      
where  $N_i^{}$  are dimensionless coupling constants  and  
\begin{eqnarray}     
\begin{array}{c}   \displaystyle   
W_{14}^{}  \;=\;  
{\rm i} \Bigl\langle \lambda\, \Bigl\{   
F_{\mu\nu}^{\rm L}+{U}^\dagger F_{\mu\nu}^{\rm R} {U} \,,\, 
L^\mu L^\nu   
\Bigr\} \Bigr\rangle   \;,  
\hspace{3em}  
W_{15}^{}  \;=\;  
{\rm i} \Bigl\langle \lambda L^\mu\, \Bigl(   
F_{\mu\nu}^{\rm L}+{U}^\dagger F_{\mu\nu}^{\rm R} {U}  
\Bigr) L^\nu \Bigr\rangle   \;,  
\vspace{2ex} \\   \displaystyle   
W_{16}^{}  \;=\;  
{\rm i} \Bigl\langle \lambda\, \Bigl\{ 
F_{\mu\nu}^{\rm L}-{U}^\dagger F_{\mu\nu}^{\rm R} {U} \,,\, L^\mu L^\nu 
\Bigr\} \Bigr\rangle   \;,  
\hspace{3em}  
W_{17}^{}  \;=\;  
{\rm i} \Bigl\langle \lambda L^\mu\, \Bigl( 
F_{\mu\nu}^{\rm L}-{U}^\dagger F_{\mu\nu}^{\rm R} {U}  
\Bigr) L^\nu \Bigr\rangle   \;.      
\end{array}      
\end{eqnarray}      
In these formulas,  
$\,\langle\cdots\rangle\equiv{\rm Tr}(\cdots),\,$    
\begin{eqnarray}   \label{fields}   
\lambda  \;\equiv\;  
\mbox{${1\over2}$} \bigl( \lambda_6^{}-{\rm i}\lambda_7^{} \bigr)   \;,  
\hspace{2em}   
F_{\mu\nu}^{\rm L}  \;=\;  F_{\mu\nu}^{\rm R}  
\;=\;  -eQ F_{\mu\nu}^{}   \;,   
\hspace{2em}
U  \;=\;  {\rm e}^{{\rm i}\phi/f_{\!\pi}^{}}   \;,  
\hspace{2em}  
L_\mu^{}  \;=\;  {\rm i} {U}^\dagger\, D_\mu^{}{U}   \;,  
\end{eqnarray}      
where  
$\,F_{\mu\nu}^{}=\partial_\mu^{}A_\nu^{}-\partial_\nu^{}A_\mu^{}\,$   
is the photon field-strength tensor, 
$\,Q={\rm diag}(2,-1,-1)/3\,$  is the quark-charge matrix, 
$\phi$  is a  $\,$3$\times$3$\,$  matrix containing 
the octet of pseudo-Goldstone bosons,  and  
$\,D_\mu^{}{U}=\partial_\mu^{}{U}+{\rm i}e [Q,U] A_\mu^{}.\,$   
Under chiral rotations, these fields transform as  
\begin{eqnarray}   
F_{\mu\nu}^{\rm L}  \;\rightarrow\;   
V_{\rm L}^{} F_{\mu\nu}^{\rm L} V_{\rm L}^\dagger   \;,  
\hspace{2em}  
F_{\mu\nu}^{\rm R}  \;\rightarrow\;   
V_{\rm R}^{} F_{\mu\nu}^{\rm R} V_{\rm R}^\dagger   \;.       
\hspace{2em}  
{U}  \;\rightarrow\;  V_{\rm R}^{} U V_{\rm L}^\dagger   \;,  
\hspace{2em}  
L_\mu^{}  \;\rightarrow\;   
V_{\rm L}^{} L_\mu^{} V_{\rm L}^\dagger   \;.       
\end{eqnarray}      
In terms of this effective Lagrangian, the~SM  contribution  
to~$F_{\rm E}^{}$  can be written as,
\begin{eqnarray}   
F_{\rm E}^{}  \;=\;  
{-m_K^3\over\sqrt{2}\, f_{\pi}^3} 
\left( N_{14}^{}-N_{15}^{}-N_{16}^{}-N_{17}^{} \right) 
\;+\;  \mbox{${\cal O}(p^4)$ loop terms}   \;.     
\end{eqnarray}      
This combination of constants ($N_i^{}$)  is not known 
from experiment so one needs to resort to model calculations 
as in Ref.~\cite{EckKW}. The loop terms, on the other hand, 
are known~\cite{DamIsi1}. 
      
The standard assumption is that the real parts of  the 
$N_i^{}$  $(i=14,\cdots,17)$ arise mostly from the 
dominant octet quark-operator $\,Q_1^{}-Q_2^{}$  
(in the notation of Ref.~\cite{BucBL}). They have been 
calculated in a couple of models in Ref.~\cite{EckKW} 
with the result,
\begin{eqnarray}   
{\rm Re} \left( N_{14}^{}-N_{15}^{}-N_{16}^{}-N_{17}^{} \right)   
\;=\;  {-f_{\pi}^2\over 2 m_\rho^2}\, k_f^{}    \;,   
\end{eqnarray}      
where  $k_f^{}=1$ corresponds to the naive factorization model 
and $k_f=1/2$ corresponds to the so-called ``weak deformation model''.  
Taking $k_f=1$ for illustration, and including the loop terms   
calculated by D'Ambrosio and Isidori~\cite{DamIsi2} (the 0.9 term), 
one obtains the estimate,
\begin{eqnarray}   
\left| {\rm Re}\, F_{\rm E}^{} \right| _{\rm SM}^{}  \;= \;  
{m_K^3\over 2 \sqrt{2}\, f_{\pi}^{} m_\rho^2}\;+\; 0.9  
\;=\;  1.7  \;.    
\end{eqnarray}      
In order to estimate $\hat\epsilon$ one also needs an estimate for the 
imaginary parts of the matrix elements, or equivalently, of the 
$N_i^{}$. Within the standard model these imaginary parts are 
expected to be dominated by the gluonic penguin operator. 
Unfortunately a detailed estimate of the bosonization of the 
penguin operator at order $p^4$ does not exist, and we have to 
resort to dimensional analysis arguments which indicate that 
$\,{\rm Im}\, F_{\rm E}^{}/{\rm Re}\, F_{\rm E}^{}$ 
is of the same order as  
$\,{\rm Im}\, A_0^{}/{\rm Re}\, A_0^{}\sim 
\sqrt{2}\, \epsilon'/\omega \sim 1\times10^{-4}.\,$  
Taking  $\,{\rm Re}\, F_{\rm E}^{}=1\,$, and assuming that there 
is no large cancellation between the two terms in Eq.~(\ref{he}),  
we arrive at the estimate 
\begin{eqnarray}   \label{|he/e|_SM'}   
\left| {\hat\epsilon\over\epsilon} \right| _{\rm SM}^{}  
\;\raisebox{-0.4ex}{$\,\stackrel{\textstyle<}{\textstyle\sim}\,$}\;  
{f_{\!\pi}^{}\over 2\sqrt{2}\, |\epsilon|\, m_K^{}}\;       
{{\rm Im}\, A_0^{}\over{\rm Re}\, A_0^{}}   
\;\sim\;  3\times 10^{-3}   \;.  
\end{eqnarray}   
This number is almost three orders of magnitude below 
the current experimental limit in Eq.~(\ref{|he/e|_x}). It 
should be clear, however, that this is no more than an 
order of magnitude estimate.

\section{New physics and short-distance  
$\,s\rightarrow d\gamma,d g\,$  transitions\label{NP}}

Physics beyond the standard model modifies the short distance 
coefficients of the two and four-quark operators that 
contribute to the~DE electric-amplitude in  
$\,K_{\rm L}^{}\rightarrow\pi^+\pi^-\gamma,\,$ and may also 
generate further operators. In this section we discuss 
the  $\,s\rightarrow d\gamma,d g\,$  operators, as they 
can be  significantly enhanced in certain 
models~\cite{ColIP,BurCIRS,sd1,sd2}.    

The effective Hamiltonian responsible for the short-distance  
$\,s\rightarrow d\gamma,d g\,$  transitions can be written, 
following the notation of  Ref.~\cite{BurCIRS},  as  
\begin{eqnarray}   \label{H_eff}
{\cal H}_{\rm eff}^{}  \;=\;  
C_{\gamma}^+ Q_{\gamma}^+ + C_{\gamma}^- Q_{\gamma}^- 
+ C_{g}^+ Q_{g}^+ + C_{g}^- Q_{g}^- 
\;+\;  {\rm H.c.}   \;,    
\end{eqnarray}      
where  $C_{\gamma,g}^\pm$  are the Wilson coefficients and  
\begin{eqnarray}   \label{Q+-}   
\begin{array}{c}   \displaystyle  
Q_{\gamma}^\pm  \;=\;    
{e Q_d^{}\over16\pi^2} \left( 
\bar{s}_{\rm L}^{}\, \sigma_{\mu\nu}^{}\, d_{\rm R}^{}  
\pm \bar{s}_{\rm R}^{}\, \sigma_{\mu\nu}^{}\, d_{\rm L}^{}  
\right) F^{\mu\nu}   \;,    
\vspace{1ex} \\   \displaystyle  
Q_{g}^\pm  \;=\;    
{g_s^{}\over16\pi^2} \left( 
\bar{s}_{\rm L}^{}\, \sigma_{\mu\nu}^{} t_a^{}\, d_{\rm R}^{}  
\pm \bar{s}_{\rm R}^{}\, \sigma_{\mu\nu}^{} t_a^{}\, d_{\rm L}^{}  
\right) G_a^{\mu\nu} 
\end{array}      
\end{eqnarray}      
are the so-called electromagnetic- and chromomagnetic-dipole 
operators, respectively, with  $\,eQ_d^{}=-e/3\,$  being 
the  $d$-quark  charge and  $G_a^{\mu\nu}$  being the gluon   
field-strength tensor.  
We notice that  $Q_{\gamma,g}^\pm$  transform as  
$\,\bigl(\bar{3}_{\rm L}^{},3_{\rm R}^{}\bigr)\pm  
\bigl(3_{\rm L}^{},\bar{3}_{\rm R}^{}\bigr)$   
under  
$\,\rm SU(3)_{L}^{}$$\times$$\rm SU(3)_{R}^{}\,$   
transformations,  and   
$Q_{\gamma,g}^{+}$  $ \bigl( Q_{\gamma,g}^{-} \bigr) $  
are even (odd) under parity.  
We further notice that  
$Q_{\gamma,g}^+$  $ \bigl( Q_{\gamma,g}^{-} \bigr) $  
are also even (odd) under  a~$CPS$  transformation 
(a $CP$  operation followed by interchanging the  $s$ and $d$  
quarks).  
These magnetic-dipole operators are expected to produce 
the most important contributions to the  
$\,s\rightarrow d\gamma,d g\,$  transitions.      
In this section we estimate the impact of these operators by   
employing a chiral-Lagrangian approach combined with naive   
dimensional-analysis~\cite{nda,weinberg} to evaluate the 
necessary hadronic matrix elements.

The chiral Lagrangian that represents the  $\,s\rightarrow d\gamma,d
g\,$  operators is constructed as follows. 
We begin by defining the objects  
\begin{eqnarray}   \label{F_LR,F_RL}
\lambda_{\rm LR}^{}  \;=\;  \lambda_{\rm RL}^{}  
\;\equiv\;  \lambda   \;,  
\end{eqnarray}      
with which we can rewrite  $Q_\gamma^\pm$  in the form    
\begin{eqnarray}   \label{Q_gamma}  
Q_{\gamma}^\pm  \;=\;    
{e Q_d^{}\over16\pi^2} \left( 
\bar{q}_{\rm L}^{}\, \lambda_{\rm LR}^{} \sigma_{\mu\nu}^{}\, 
q_{\rm R}^{}  
\pm \bar{q}_{\rm R}^{}\, \lambda_{\rm RL}^{} 
\sigma_{\mu\nu}^{}\, q_{\rm L}^{}  
\right) F^{\mu\nu}   \;,    
\end{eqnarray}      
where  $\,q=(u$ $d$ $s)^{\rm T}.\,$        
If we imagine that under a chiral rotation  
\begin{eqnarray}   \label{F->F'}
\lambda_{\rm LR}^{}  \;\rightarrow\;  
V_{\rm L}^{} \lambda_{\rm LR}^{} V_{\rm R}^\dagger   \;,
\hspace{2em}  
\lambda_{\rm RL}^{}  \;\rightarrow\;  
V_{\rm R}^{} \lambda_{\rm RL}^{} V_{\rm L}^\dagger   \;,
\end{eqnarray}      
where  $\,V_{\rm L,R}\in\rm SU(3)_{L,R}^{},\,$  
then the quark operators inside the brackets in  
Eq.~(\ref{Q_gamma})  would be chirally invariant.   
Consequently, we construct the effective Lagrangian 
by writing chiral invariant terms with one power of 
$\lambda_{\rm LR}^{}$  or  $\lambda_{\rm RL}^{}$ 
assumed to transform as in  Eq.~(\ref{F->F'}), and with an 
explicit photon field strength tensor $F^{\mu\nu}$.  
We also require the effective Lagrangian to have the 
same  parity and  $CPS$  structure  of~$Q_{\gamma}^\pm$.
In this regard,  $\lambda_{\rm LR}^{}$  and  
$\lambda_{\rm RL}^{}$  interchange places under parity, 
but remain unchanged under charge conjugation.

There are many possible chiral realizations of  $Q_{\gamma}^\pm$.
As relevant examples at leading  order,~$p^4$,  we write down  
\begin{eqnarray}   \label{L^gamma,eff}
{\cal L}_\gamma^{(\pm)}  \;=\;  
{\rm i}\beta_\gamma^{\pm} e Q_d^{}\, \left\langle   
\lambda_{\rm LR}^{} {U} L_\mu^{} L_\nu^{}   
\pm   
\lambda_{\rm RL}^{} L_\mu^{} L_\nu^{} {U}^\dagger 
\right\rangle F^{\mu\nu}  
\;+\;  {\rm H.c.}   \;,   
\end{eqnarray}      
where  ${\cal L}_\gamma^{(\pm)}$  have the same symmetry   
properties as those of  $Q_{\gamma}^{\pm}$.  
Employing naive dimensional-analysis~\cite{nda,weinberg}, 
we obtain the order-of-magnitude estimate   
\begin{eqnarray}   \label{beta,nda}   
\beta_\gamma^{\pm}  \;=\;  
{C_{\gamma}^{\pm} f_{\!\pi}^{}\over 16\pi^2}\,  
{f_{\!\pi}^{}\over\Lambda}   \;,    
\end{eqnarray}      
where  $\,\Lambda=4\pi f_{\!\pi}^{}.\,$   
Notice that  ${\cal L}_\gamma^{(\pm)}$  are not suppressed by   
light-quark masses, as is appropriate for the new-physics  
interactions of interest in Eq.~(\ref{H_eff}).   
As was noted before, within the SM the short-distance operator 
is suppressed by light-quark masses,  and this results in 
a different chiral Lagrangian (of  order~$p^6$  at least) 
involving the usual chiral-symmetry breaking 
factor~\cite{DamIsi2}.

The contribution of the  $\,s\rightarrow d\gamma$  
operators  to the direct-emission electric-amplitude of 
$\,K^0\rightarrow\pi^+\pi^-\gamma\,$  comes from  
${\cal L}_\gamma^{(-)}$  and is given by  
\begin{eqnarray}   
\left( F_{\rm E}^{} \right) _\gamma^{}  \;=\;  
{\sqrt{2}\, \beta_{-}^{}\over 3G_8^{}f_{\!\pi}^2}\, 
{m_K^3\over f_{\!\pi}^3}   
\;=\;  
{m_K^3\over 96\sqrt{2}\, \pi^3\, G_8^{}f_{\!\pi}^4}\; C_\gamma^-   \;.  
\end{eqnarray}   
Then, if the new-physics contribution to  $F_{\rm E}^{}$  
is such that  
$\,{\rm Im}\, F_{\rm E}^{}/{\rm Re}\, F_{\rm E}^{}\gg
{\rm Im}\, A_0^{}/{\rm Re}\, A_0^{}\,$  (i.e. $CP$ violation is 
dominated by new physics) 
in  $\hat\epsilon$,  the contribution of  $C_\gamma^-$  
to  $\,\hat\epsilon/\epsilon\,$  is  
\begin{eqnarray}   \label{|he/e|_gamma} 
\left| {\hat\epsilon\over\epsilon} \right| _\gamma^{}  
\;\simeq\;  
{ m_K^4\; \left| {\rm Im}\, C_\gamma^- \right|  \over  
 384\sqrt{2}\, \pi^3\, |\epsilon| f_{\!\pi}^2\, {\rm Re}\, A_0^{}}  
\;\simeq\;  
6.9\times10^5\; {\rm GeV}\; 
\left| {\rm Im}\, C_\gamma^- \right|  \;.  
\end{eqnarray}   
This and Eq.~(\ref{|he/e|_x})  imply the current limit  
\begin{eqnarray}   \label{ImC_gamma,limit}    
\left| {\rm Im}\, C_\gamma^{-} \right|      
\;<\;  2.7\times 10^{-6}\; {\rm GeV}^{-1}   \;.    
\end{eqnarray}   
Future measurements~\cite{eta+-g3}  may improve this value by 
about a factor of 10.

Turning now to the chiral realization of 
the  $\,s\rightarrow d g\,$  operators  $Q_{g}^\pm$,  
we rewrite them as      
\begin{eqnarray}   \label{Q_gluon}  
Q_{g}^\pm  \;=\;    
{g_{\rm s}^{}\over16\pi^2} \left( 
\bar{q}_{\rm L}^{}\, \lambda_{\rm LR}^{} \sigma_{\mu\nu}^{}\,     
t_a^{} G_a^{\mu\nu}\, q_{\rm R}^{}  
\pm   
\bar{q}_{\rm R}^{}\, \lambda_{\rm RL}^{} \sigma_{\mu\nu}^{}\,    
t_a^{} G_a^{\mu\nu}\, q_{\rm L}^{}  
\right)   \;.      
\end{eqnarray}      
Since  $Q_{g}^\pm$  do not contain the photon field,  
in constructing the corresponding chiral Lagrangian involving  
an  $F_{\mu\nu}^{}$,  we employ the chiral field-strength 
tensors  $F_{\mu\nu}^{\rm L}$  and  $F_{\mu\nu}^{\rm R}$,  
which were defined in  Eq.~(\ref{fields}).   
These tensors transform under parity as  
$\,F_{\rm L,R}^{\mu\nu}\rightarrow 
\bigl( F_{\rm R,L}^{} \bigr) _{\mu\nu}^{}\,$  
and under charge conjugation as  
$\,F_{\rm L,R}^{\mu\nu}\rightarrow 
- \bigl( F_{\rm R,L}^{\rm T} \bigr) ^{\mu\nu}.\,$

Thus, examples of chiral Lagrangians at  order~$p^4$  that 
correspond to  $Q_{g}^\pm$ are   
\begin{eqnarray}   \label{L^gluon,eff}
{\cal L}_g^{(\pm)}  \;=\;  
{\rm i}\beta_g^{\pm}\, \left\langle 
\lambda_{\rm LR}^{} {U} L^\mu \left(   
F_{\mu\nu}^{\rm L}+{U}^\dagger F_{\mu\nu}^{\rm R} {U} 
\right) L^\nu   
\pm  
\lambda_{\rm RL}^{} L^\mu\, \left(    
F_{\mu\nu}^{\rm L}+{U}^\dagger F_{\mu\nu}^{\rm R} {U}    
\right) L^\nu {U}^\dagger   
\right\rangle   
\;+\;  {\rm H.c.}   \;,   
\end{eqnarray}      
where, from naive dimensional-analysis~\cite{nda,weinberg}, 
\begin{eqnarray}   \label{beta_gluon,nda}   
\beta_g^{\pm}  \;=\;  
{C_g^{\pm} f_{\!\pi}^{}\over 16\pi^2}\,  
{f_{\!\pi}^{}\over\Lambda}\, g_{\rm s}^{}   \;.   
\end{eqnarray}      
For numerical estimates we use $\,g_{\rm s}^{}\sim\sqrt{4\pi},\,$  
corresponding  to a strong coupling  $\,\alpha_{\rm s}^{}\sim1.\,$   
The contribution to  $F_{\rm E}^{}$  arises from  
${\cal L}_g^{(-)}$  and is given by  
\begin{eqnarray}   
\left( F_{\rm E}^{} \right) _g^{}  \;=\;  
{ g_{\rm s}^{} m_K^3  \over   
 96\sqrt{2}\, \pi^3\, G_8^{}f_{\!\pi}^4 }\; C_g^-   \;.  
\end{eqnarray}   
We again assume that $CP$ violation is dominated by the  
new physics, in such a way that  
$\,{\rm Im}F_{\rm E}^{}/{\rm Re}F_{\rm E}^{}$ 
$\gg {\rm Im}\, A_0^{}/{\rm Re}\, A_0^{},\,$  to obtain
\begin{eqnarray}   \label{|he/e|_gluon} 
\left| {\hat\epsilon\over\epsilon} \right| _g^{}  
\;\simeq\;  
{ g_{\rm s}^{}\, m_K^4\; \left| {\rm Im}\, C_g^- \right|  \over  
 384\sqrt{2}\, \pi^3\, |\epsilon|\, f_{\!\pi}^2\, 
 {\rm Re}\, A_0^{} }  
\;\simeq\;  
2.4\times10^6\; {\rm GeV}\; \left| {\rm Im}\, C_g^- \right|  \;,    
\end{eqnarray}   
which, with  Eq.~(\ref{|he/e|_x}),  implies the limit  
\begin{eqnarray}   \label{ImC_gluon,limit}    
\left| {\rm Im}\, C_g^{-} \right|      
\;<\;  7.8\times 10^{-7}\; {\rm GeV}^{-1}   \;.    
\end{eqnarray}   

Since it is known~\cite{BurCIRS,DonHol,BerFG} that  $Q_{g}^-$  
also contributes to  $\epsilon'$,  we can use the measured 
value of  $\epsilon'$  to derive another limit on 
the contribution of  $Q_{g}^-$  to  $\hat\epsilon$.     
The leading chiral realization of  $Q_{g}^-$  that 
contributes to the  $\,K_{\rm L}^{}\rightarrow\pi\pi\,$   
amplitude is of  order~$p^2$,  as the  ${\cal O}(p^0)$  
realization does not contribute once tadpole diagrams 
have been properly taken into account~\cite{DonHol}.   
An example that we can construct is\footnote{A similar 
Lagrangian has been given in  Ref.~\cite{HeVal2}.}     
\begin{eqnarray}   
{\cal L}_g^{(2)}  \;=\;  
\gamma_g^{-} f_{\!\pi}^2\, 
\left\langle \lambda \left( {U}-{U}^\dagger \right) \right\rangle 
\left\langle L^\mu L_\mu^{} \right\rangle   
\;+\;  {\rm H.c.}   \;,  
\end{eqnarray}   
where  $\gamma_g^-$  is a dimensionless coupling constant.   
Using the naive dimensional-analysis prescribed in 
Ref.~\cite{weinberg},  we find the order-of-magnitude estimate    
\begin{eqnarray}   
\gamma_g^{-}  \;=\;  
{C_g^-\, f_{\!\pi}^{}\, g_{\rm s}^{}\over 16\pi^2}   \;.    
\end{eqnarray}   
The resulting amplitude is  
\begin{eqnarray}   
\left( A_0^{} \right) _g^{}  \;=\;  
{ C_g^-\, g_{\rm s}^{}\over 2\sqrt{2}\, \pi^2} 
\left( m_K^2-2m_\pi^2 \right)   \;,  
\end{eqnarray}      
leading to  
\begin{eqnarray}   \label{|e'/e|_gluon}   
\left| {\epsilon'\over\epsilon} \right| _g^{}   \;=\;  
{\omega\over |\epsilon|\, {\rm Re}\, A_0^{}}\;   
{g_{\rm s}^{}\over 4\pi^2} \left( m_K^2-2m_\pi^2 \right)  
\left| {\rm Im}\, C_g^- \right| \,
\;\simeq\;  
1.4\times 10^6\;  \left| {\rm Im}\, C_g^- \right|   \;.    
\end{eqnarray}      
Assuming that the current value  
$\,|\epsilon'/\epsilon|\sim 2\times10^{-3}\,$  
is saturated by the  $Q_{g}^{-}$  contribution yields
\begin{eqnarray}     
\left| {\rm Im}\, C_g^- \right| 
\;<\;  1.6\times 10^{-9}   \;,    
\end{eqnarray}   
which is a much better constraint than  
Eq.~(\ref{ImC_gluon,limit}).   
Furthermore, combining Eqs.~(\ref{|e'/e|_gluon})  
and~(\ref{|he/e|_gluon})  results in  
\begin{eqnarray}   \label{|he/e|_gluon,limit} 
\left| {\hat\epsilon\over\epsilon} \right| _g^{}  \;\simeq\;  
{ m_K^4  \over  96\sqrt{2}\, \pi\, \omega e\, f_{\!\pi}^2\, 
 \left( m_K^2-2m_\pi^2 \right) }  
\left| {\epsilon'\over\epsilon} \right| _g^{}   
\;\simeq\;  
1.8\; 
\left| {\epsilon'\over\epsilon} \right| _g^{}   
\;<\;  4\times 10^{-3}   \;.  
\end{eqnarray}   
This is twice as large as the standard model result, 
but the two should be considered equivalent within the 
uncertainties of our estimates. 
From this we conclude that improved measurements of 
$\hat\epsilon$ are more important to place bounds on 
$Q_\gamma^{-}$. Bounds on $Q_g^{-}$ from $\hat\epsilon$ 
are not likely to be competitive with bounds from 
$\epsilon'/\epsilon$ in the foreseeable future. 
We now turn our attention to two specific models for 
$Q_\gamma^{-}$.

\subsection{Left-right symmetric models\label{LRSM}}   
   
In left-right symmetric models the coefficients of  
$Q_\gamma^{-}$ and $Q_g^{-}$ can be enhanced because 
the mixing of left- and right-handed  $W$-bosons removes the 
helicity suppression present in the standard model.    
Variations of this model have been studied in the context of  
$\,b\rightarrow s\gamma\,$  in detail~\cite{FujYam,bsg2}.  
  
We start from the effective Lagrangian that results 
from integrating out the heavy right-handed  $W$. 
This can be written down directly by following the formalism of  
Ref.~\cite{PecZha}. In the unitary gauge, 
\begin{eqnarray}   
{\cal L}_{\rm RH}^{}  \;=\;  
-{g_2^{}\over\sqrt{2}} 
\left( \bar{u}_{\rm R}^{} \hspace{1em} \bar{c}_{\rm R}^{} 
      \hspace{1em} \bar{t}_{\rm R}^{} \right) \, \gamma^\mu\, 
\tilde{V} 
\left( \begin{array}{c}   \displaystyle  
d_{\rm R}^{}   \vspace{1ex} \\   \displaystyle   
s_{\rm R}^{}   \vspace{1ex} \\   \displaystyle  
b_{\rm R}^{}  
\end{array} \right) 
W_\mu^+    
\;+\;  {\rm h.c.}   \;,  
\end{eqnarray}      
where  $\tilde{V}$  is a $3\times3$  unitary matrix  having 
elements  
$\,\tilde{V}_{qq'}^{}=V_{qq'}^{}\kappa_{qq'}^{\rm R},\,$  
with  $V_{qq'}^{}$  being  CKM-matrix elements and  
$\kappa_{qq'}^{\rm R}$  complex numbers.   
In writing  ${\cal L}_{\rm eff}^{}$  above, we have ignored 
modifications to the left-handed $W$-couplings which do not 
lead to enhanced effects. Using the results of 
Refs.~\cite{FujYam,AbdVal} we write,
\begin{eqnarray}     
\begin{array}{c}   \displaystyle   
C_{\gamma,\rm RH}^{-} (m_W^{})  \;=\;   
{G_{\rm F}^{}\over\sqrt{2}}   
\sum_{q=c,t} V_{qd}^{} V_{qs}^{*}\,   
\left( \kappa_{qs}^{\rm R*}-\kappa_{qd}^{\rm R} \right) \, 
m_q^{}\,    
{F_{\rm RH}^{}(x_q^{})\over Q_d^{}}   \;,   
\vspace{2ex} \\   \displaystyle   
C_{g,\rm RH}^{-} (m_W^{})  \;=\;   
{G_{\rm F}^{}\over\sqrt{2}}   
\sum_{q=c,t} V_{qd}^{} V_{qs}^{*}\,     
\left( \kappa_{qs}^{\rm R*}-\kappa_{qd}^{\rm R} \right) \, 
m_q^{}\, G_{\rm RH}^{}(x_q^{})   \;,  
\end{array}      
\end{eqnarray}      
where  $\,x_q^{}=m_q^2/m_W^2\,$  and     
\begin{eqnarray}   
F_{\rm RH}^{} (x)  \;=\;   
{-3x^2+2x\over(x-1)^3} \, \ln x-{5x^2-31x+20\over 6(x-1)^2}   \;,    
\hspace{2em}   
G_{\rm RH}^{} (x)  \;=\;   
{6x\, \ln x\over(x-1)^3} - {3+3x\over (x-1)^2} - 1   \;.    
\end{eqnarray}      
For our numerical estimates, we will use  
$\,\alpha_s^{}(m_Z^{})=0.119,\,$  
$\,m_c^{}=1.25\,\rm GeV,\,$  $\,m_t^{}=173.8\,\rm GeV,\,$  
and  the CKM-matrix elements in 
the Wolfenstein parameterization from Ref.~\cite{mele}: 
$\,\lambda=0.22,\,$  $\,A=0.82,\,$  $\,\rho=0.16,\,$  and  $\,\eta=0.38.\,$  
This gives,
\begin{eqnarray}     
C_{\gamma,\rm RH}^{-}  &\simeq& 
\left[    
1\times 10^3\; V_{cd}^{} V_{cs}^*
 \left( \kappa_{cs}^{\rm R*}-\kappa_{cd}^{\rm R} \right)   
+ 7\times 10^4\; V_{td}^{} V_{ts}^* 
 \left( \kappa_{ts}^{\rm R*}-\kappa_{td}^{\rm R} \right)   
\right] \times 10^{-7} \;{\rm GeV}^{-1}   \;,  
\end{eqnarray}      
\begin{eqnarray}     
C_{g,\rm RH}^{-}  &\simeq&  
\left[    
-4\times 10^2\; V_{cd}^{} V_{cs}^*  
\left( \kappa_{cs}^{\rm R*}-\kappa_{cd}^{\rm R} \right)   
- 2\times 10^4\; V_{td}^{} V_{ts}^* 
 \left( \kappa_{ts}^{\rm R*}-\kappa_{td}^{\rm R} \right)   
\right] \times 10^{-7} \;{\rm GeV}^{-1}   \;.  
\end{eqnarray}      
It then follows that  
\begin{eqnarray}   
\left| {\hat\epsilon\over\epsilon} \right| _{\gamma,\rm RH}^{} 
\;\simeq\;  
\left| 
15\; {\rm Im} 
 \left( \kappa_{cs}^{\rm R*}-\kappa_{cd}^{\rm R} \right)   
+ {\rm Im} \left[ 
 (1.3 - 0.6\, {\rm i}) 
 \left( \kappa_{ts}^{\rm R*}-\kappa_{td}^{\rm R} \right) \right]    
\right|   \;. 
\end{eqnarray}   
These same couplings contribute to $\epsilon'/\epsilon$ in this 
model through $Q_g$,
\begin{eqnarray}   
\left| {\epsilon'\over\epsilon} \right| _{g,\rm RH}^{}  
\;\simeq\;  
 \left| 
12\; {\rm Im} 
\left( \kappa_{cs}^{\rm R*}-\kappa_{cd}^{\rm R} \right)   
+ {\rm Im} \left[ 
 (0.8 - 0.4\, {\rm i})  
 \left( \kappa_{ts}^{\rm R*}-\kappa_{td}^{\rm R} \right) \right]    
\right|   \;.    
\end{eqnarray}      
From these results, we see that the contribution of $Q^-_\gamma$ 
to $\hat\epsilon$ is also constrained by the contribution of 
$Q^-_g$ to $\epsilon'$.  
For example, if all 
$\,{\rm Im} \bigl( \kappa_{qs}^{\rm R*}-\kappa_{qd}^{\rm R} \bigr) \,$  
are of the same order, then the  $c$-quark  contributions 
dominate both observables and  
$\,|\hat\epsilon|\sim 1.3\, |\epsilon'|\,$.  
Similarly, if the $t$-quark intermediate-state dominates, 
$\,|\hat\epsilon|\sim 1.7\, |\epsilon'|\,$.  
If one includes perturbative-QCD running of the Wilson 
coefficients from  the $W$-mass scale down to a scale near 
the charm-quark  mass~\cite{BucBL,running}, these numbers 
slightly change to  $\,|\hat\epsilon|\sim 1.5\, |\epsilon'|$   
and  $1.9\, |\epsilon'|,\,$  respectively.  
With $|\epsilon'/\epsilon|_{g,{\rm RH}}\sim 2 \times 10^{-3}$, 
this results in
\begin{eqnarray}   
\left| {\hat\epsilon\over\epsilon} \right| _{\gamma,\rm RH}^{} 
\;\sim\;  4\times 10^{-3}   \;,     
\end{eqnarray}   
which is comparable to the contribution of  $Q_g^{}$  
as in Eq.~(\ref{|he/e|_gluon,limit}).    
In full generality, $\hat\epsilon/\epsilon$  and $\epsilon'/\epsilon$ 
are proportional to different combinations of the 
$\kappa$'s, so that it is possible for $\hat\epsilon$ to be 
significantly larger than $\epsilon'$, although this does not seem 
likely.

\subsection{Supersymmetric Models\label{susy}}

In certain supersymmetric models, one can generate 
the  $\,s\rightarrow d\gamma\,$  operators at one-loop 
via intermediate squarks and gluinos resulting in large
$C_{\gamma,g}^{}$.  
The enhancement is due both to the strong coupling constant and 
to the removal of chirality suppression present in the standard 
model. We follow  Ref.~\cite{GabGMS} 
and work in the so-called mass-insertion approximation.  
The full expressions for  $C_{\gamma,\rm susy}^{-}$  can be 
found in  Ref.~\cite{GabGMS}.    
Here we are interested only in the terms enhanced by 
$m_{\tilde{g}}^{}/m_s^{}$,  with $m_{\tilde{g}}^{}$  being 
the average gluino-mass, and they are
\begin{eqnarray}     
\begin{array}{c}   \displaystyle   
C_{\gamma,\rm susy}^{-} (m_{\tilde{g}}^{})  \;=\;   
{\pi\, \alpha_{\rm s}^{}(m_{\tilde{g}}^{})\over m_{\tilde{g}}^{}}
\left[ \bigl( \delta_{21}^{d} \bigr) _{\rm LR}^{} 
      - \bigl( \delta_{21}^{d} \bigr) _{\rm RL}^{} \right] 
F_{\rm susy}^{}(x_{gq}^{})   \;,  
\vspace{2ex} \\   \displaystyle   
C_{g,\rm susy}^{-} (m_{\tilde{g}}^{})  \;=\;   
{\pi\, \alpha_{\rm s}^{}(m_{\tilde{g}}^{})\over m_{\tilde{g}}^{}}
\left[ \bigl( \delta_{21}^{d} \bigr) _{\rm LR}^{} 
      - \bigl( \delta_{21}^{d} \bigr) _{\rm RL}^{} \right] 
G_{\rm susy}^{}(x_{gq}^{})   \;,  
\end{array}      
\end{eqnarray}      
where  the  $\delta$'s  are the parameters of the mass-insertion 
formalism,  $\,x_{gq}^{}=m_{\tilde{g}}^2/m_{\tilde{q}}^2,\,$  
with  $m_{\tilde{q}}$  being the average squark-mass,  and  
\begin{eqnarray}   
\begin{array}{c}   \displaystyle   
F_{\rm susy}^{} (x)  \;=\;   
{ 4x\, \bigl( 1+4x-5x^2+4x\,\ln x+2x^2\,\ln x \bigr)  \over  
 3(1-x)^4 }   \;,    
\vspace{2ex} \\   \displaystyle   
G_{\rm susy}^{} (x)  \;=\;   
{ x\, \bigl( 22-20x-2x^2-x^2\,\ln x+16x\,\ln x+9\,\ln x \bigr)  
 \over  3(1-x)^4 }   \;.    
\end{array}      
\end{eqnarray}      
For our estimates, it is sufficient to approximate 
$\,F_{\rm susy}^{}(x) \sim F_{\rm susy}^{}(1)=2/9\,$  
and  
$\,G_{\rm susy}^{}(x) \sim G_{\rm susy}^{}(1)=-5/18\,$.  
This approximation introduces an error smaller than a factor 
of two.

Then, by means of  Eqs.~(\ref{|he/e|_gamma})  
and~(\ref{|e'/e|_gluon}),  we obtain  
\begin{eqnarray}     
\left| {\hat\epsilon\over\epsilon} \right| _{\gamma,\rm susy}^{}     
&\simeq&  
92\,   
\left( { \alpha_{\rm s}^{}(m_{\tilde{g}}^{})  \over   
        \alpha_{\rm s}^{}(500\,{\rm GeV}) } \right) \,   
{500\,{\rm GeV}\over m_{\tilde{g}}^{}}\,   
\left| {\rm Im}   
\left[ \bigl( \delta_{12}^{d} \bigr) _{\rm LR}^{}   
      - \bigl( \delta_{12}^{d} \bigr) _{\rm RL}^{} \right]   
\right|  
\nonumber \\    
&\simeq&    
{0.4}\;\left| {\epsilon'\over\epsilon} \right| _{g,\rm susy}^{}   \;.       
\end{eqnarray}   
We find that the $Q^-_g$ contribution to 
$\epsilon'$ constrains the $Q^-_\gamma$ contribution to 
$\hat\epsilon$ as it also happened in the left-right model. 
Once again with  
$|\epsilon'/\epsilon|_{g,{\rm susy}}\sim 2\times10^{-3},\,$  
the contribution of  $C_{\gamma,\rm susy}^{}$ to $\hat\epsilon$  
can reach  
\begin{eqnarray}     
\left| {\hat\epsilon\over\epsilon} \right| _{\gamma,\rm susy}^{}    
\;\sim\;  8 \times 10^{-4}   \;.    
\end{eqnarray}   
If we incorporate the running of the coefficients down to 
a low-energy scale~\cite{BurCIRS}, we will instead have 
$\,|\hat\epsilon/\epsilon| _{\gamma,\rm susy}^{}    
\sim 1.7\times 10^{-3},\,$  
which is roughly comparable to the contribution of  $Q_g^{}$  
in  Eq.~(\ref{|he/e|_gluon,limit}).    
Our estimate, translated into  $\tilde{\epsilon}_{+-\gamma}'$,  
is three times smaller than the estimate of this parameter in  
Ref.~\cite{ColIP}.  
The factor of three can be traced back to the use in 
Ref.~\cite{ColIP} of a hadronic matrix element three times larger 
than ours. 
These differences are a good indication of the level of 
precision that can be expected from this type of estimates.

\section{Summary and Conclusions}

We have reanalyzed direct  $CP$  violation in the electric 
amplitude for the decay    
$\,K_{\rm L}^{}\rightarrow\pi^+\pi^-\gamma.\,$  The 
dominant $CP$ violating observable arises from an 
interference between the IB (inner bremsstrahlung) 
and the E1 direct emission amplitudes. The previous theoretical 
definition of $\eta_{+-\gamma}$ results in a quantity that 
varies with kinematic variables and that differs from the 
previously used experimental definition of $\eta_{+-\gamma}$. 
To clarify this situation we have introduced a new theoretical quantity,  
$\hat\epsilon$, to parameterize {\it new} direct $CP$ violation 
in this decay. This quantity is a constant, and by simple 
dimensional analysis is expected to be between a few and five 
times larger than $\epsilon'$. 

The experimental observable 
$\epsilon'_{+-\gamma}$ is related to $\hat\epsilon$ by 
a normalized integration over phase space. We find that the 
very different $E_\gamma^*$ dependence of the IB and E1-DE 
amplitudes introduces a large kinematic suppression into 
$\epsilon'_{+-\gamma}$. In particular, for $E_\gamma^* > 20$~MeV, 
$|\epsilon'_{+-\gamma}|=0.041|\hat\epsilon|$, and thus much 
smaller than $\epsilon'$. This kinematic suppression can be reduced 
by increasing the minimum $E_\gamma^*$ accepted at the cost of 
diminished statistics. 

From presently available data we have extracted the bound 
$|\hat\epsilon/\epsilon|< 1.9$, significantly 
larger than the theoretical expectation 
$|\hat\epsilon/\epsilon| < 0.01$.

We have estimated the value of $\hat\epsilon$ in several models. 
The estimates are hindered by unknown hadronic matrix elements 
and must be considered order of magnitude estimates. 
The general conclusions from these estimates are

\begin{itemize}

\item If the $CP$-violating phase of the E1-DE amplitude is 
similar in size to the $CP$ violating phase of $A_0$, and the 
two do not cancel, then $\hat\epsilon$ is about as large 
as $\epsilon'$ and could be as much as five times larger. 
This is a refined form of the naive dimensional analysis estimate. 
This is the case in the standard model estimate of 
$\hat\epsilon$ where the $CP$-violating phase cannot be 
computed and is simply assumed to be of the same order as 
the phase of $A_0$.

\item Beyond the standard model, in models where the 
chromomagnetic dipole operator is enhanced, the maximum 
value of $\hat\epsilon$ is directly constrained by the 
fraction of $\epsilon'$ that is attributed to the new physics. 
Without invoking a fine-tuned cancellation in $\epsilon'$, 
this implies that in these models $\hat\epsilon$ is also 
at most a few to five times larger than $\epsilon'$. This 
happens, of course, because the bound from $\epsilon'$ is 
equivalent to the simple assumption that the $CP$ violating 
phases of the E1-DE amplitude and $A_0$ are of similar size.

\item In principle, $\hat\epsilon$ could be 
larger in models in which it is induced primarily by an 
electromagnetic dipole operator which does not contribute 
significantly to $\epsilon'$. In the specific models that 
we studied, however, the coefficients of the electromagnetic 
and chromomagnetic dipole operators are highly correlated. 
For this reason, in these models, the maximum $\hat\epsilon$ 
is, once again, limited by the fraction of $\epsilon'$ that 
we attribute to the new physics.

\end{itemize}

To conclude, $\hat\epsilon$ is expected to be comparable to 
$\epsilon'$ although it can be as much as five times larger 
both in the standard model and beyond. Unfortunately, the 
experimental observable $\epsilon'_{+-\gamma}$ is kinematically 
suppressed, 
calculated to be  
$\,\sim 0.04-0.09 \hat\epsilon\,$  for  
$\,E_\gamma^*>20-50$~MeV.

\bigskip\bigskip  
   
\noindent 
{\bf Acknowledgments}$\;$    
This work  was supported in part by DOE under contract number 
DE-FG02-92ER40730. 
We thank the Brookhaven Theory Group, the Fermilab Theory Group, 
and the Special Research Centre for the Subatomic 
Structure of Matter at the University of Adelaide 
for their hospitality and partial support while this 
work was completed.   
We are grateful to J.~Belz and Y.B.~Hsiung for conversations   
and for discussions of KTeV data.

\end{document}